%% file: draft3.tex
\documentclass[letterpaper,11pt]{article}
\usepackage{amsmath, amsthm, amssymb}
\usepackage[usenames, dvipsnames]{color}
\usepackage[normalem]{ulem} 
\usepackage{fullpage}
\usepackage{cite}
\usepackage[numbers, sort&compress]{natbib}
\usepackage[noend]{algorithmic}
\usepackage{tikz, pgfplots}
\usepackage{enumerate}
\usepackage{hyperref}
\usepackage{xspace,color}
\usepackage{graphicx}
\usepackage{caption}
\usepackage{subcaption}
\usepackage{enumitem,linegoal}
\usepackage[linesnumbered,ruled,vlined]{algorithm2e}
\usepackage{comment}	
\usepackage{ctable}					
\newtheorem{claim}{Claim}[section]

\newtheorem{observation}{Observation}[section]

\usepackage{mathtools}
\usepackage[flushleft]{threeparttable}
\usepackage{verbatim}
\usepackage{setspace}
\usepackage{bbm}
\usepackage{thm-restate}
\usepackage{footnote}
\usepackage{cleveref}
\crefname{LP}{LP}{LPs}
\crefname{theorem}{Theorem}{Theorems}
\crefname{lemma}{Lemma}{Lemmas}
\crefname{section}{Section}{Sections}
\crefname{subsection}{Section}{Sections}
\crefname{definition}{Definition}{Definitions}
\crefname{figure}{Figure}{Figures}
\crefname{table}{Table}{Tables}
\crefformat{equation}{(#2#1#3)}
\crefname{claim}{Claim}{Claims}
\crefname{observation}{Observation}{Observations}
\crefname{example}{Example}{Examples}
\usepackage{tikz}
\usetikzlibrary{positioning, decorations.pathmorphing}
\usepackage{pgfplots}
\pgfplotsset{compat=1.18}
\usepackage{hyperref}
\usepackage{tikz}
\usetikzlibrary{decorations.pathreplacing}

\makesavenoteenv{tabular}
\makesavenoteenv{table}

\usepackage[margin=1in]{geometry}

\definecolor{darkbrown}{rgb}{0.55,0.27,0.07}
\definecolor{crimsonglory}{rgb}{0,0,0}

\newtheorem{example}{Example}

\newcommand\Tstrut{\rule{0pt}{2.6ex}} 
\newcommand\Bstrut{\rule[-1.1ex]{0pt}{0pt}} 
\newcommand{\TBstrut}{\Tstrut\Bstrut} 

 \newtheorem{theorem}{Theorem}[section]
 
 \newtheorem{lemma}[theorem]{Lemma}

 \newtheorem{definition}[theorem]{Definition}

\makeatletter
\def\GrabProofArgument[#1]{ #1: \egroup\ignorespaces}
\def\proof{\noindent\textbf\bgroup Proof%
	\@ifnextchar[{\GrabProofArgument}{. \egroup\ignorespaces}}

\makeatother

\usetikzlibrary{arrows,shapes,snakes,automata,backgrounds,petri,calc}
\usepackage[latin1]{inputenc}

\input{src/macros.tex}

\newcounter{proccnt}

\newcommand{\konote}[1]{}

\usepackage[normalem]{ulem} 

\title{Beating Metric Distortion 3 by Randomizing\\ over Two Candidates}

\author{
 Masoud Seddighin\thanks{Tehran Institute for Advanced Studies.}
 \and
 Saeed Seddighin\thanks{Independent Researcher.}
}

\SetCommentSty{mycommfont}

\SetKwInput{KwInput}{Input}                
\SetKwInput{KwOutput}{Output}  
\begin{document}
	\newcommand{\ignore}[1]{}
\renewcommand{\theenumi}{(\roman{enumi}).}
\renewcommand{\labelenumi}{\theenumi}
\sloppy
\date{}

\maketitle

\thispagestyle{empty}

\begin{abstract}
\input{src/abstract.tex}
\end{abstract}
\input{src/intro.tex}
\input{src/preliminaries.tex}
\input{src/results.tex}
\input{src/building.tex}
\input{src/veto.tex}

\input{src/repair.tex}
\input{src/profiles.tex}
\input{src/pair.tex}
\input{src/blocks.tex}
\input{src/coupled.tex}
\input{src/descent3.tex}

\input{src/computation.tex}
\input{src/conclusion.tex}
\input{src/ai.tex}
\bibliographystyle{unsrtnat}	
\bibliography{draft}
\appendix

\end{document}

%% file: src/macros.tex
\newcounter{claudecnt}

\newif\ifshowclaudeedits
\showclaudeeditstrue 

\newcommand{\SC}{\operatorname{\mathsf{SC}}}          
\newcommand{\OPT}{\mathsf{OPT}}        
\newcommand{\dist}{\operatorname{dist}}      
\newcommand{\plu}{\operatorname{plu}}        
\newcommand{\defi}{\operatorname{def}}       
\newcommand{\tp}{\operatorname{top}}         
\newcommand{\E}{\mathbb{E}}
\newcommand{\R}{\mathbb{R}}
\newcommand{\pos}[1]{\left[#1\right]_+}
\newcommand{\eps}{\varepsilon}
\newcommand{\cD}{\mathcal{D}}               
\newcommand{\cP}{\mathcal{P}}               
\newcommand{\hdelta}{\hat{\delta}}
\newcommand{\tbar}{\bar{t}}

%% file: src/abstract.tex
Every deterministic voting rule has metric distortion at least $3$, while randomized rules can do better. Cai, Gao, Ramakrishnan, and Wang (EC'26) conjectured that a uniform lottery over just two candidates suffices to beat $3$. We prove that randomizing over two candidates, with possibly unequal probabilities, indeed breaks this barrier: in every election, some lottery supported on two candidates has distortion at most $3$ minus an absolute positive constant.\footnote{The bound is $3-\eps_*$, where $\eps_*$ is defined in \cref{thm:main}; in particular, $\log^*(1/\eps_*)\ge 4005$, where $\log^*$ denotes the iterated natural logarithm.} Such a lottery can be computed in polynomial time.

%% file: src/intro.tex
\section{Introduction}

Elections are among the most basic mechanisms for collective decision making. In a typical election, every voter submits a ranking of the candidates and a voting rule aggregates these rankings into a winner. A natural way to evaluate a voting rule is to imagine that every voter has a cardinal cost for every candidate, and to ask how much the total cost of the elected candidate can exceed the optimum. Since the rule only observes rankings, it cannot always find the optimal candidate. The worst-case ratio between the cost of its outcome and the optimal cost is called the \emph{distortion} of the rule~\cite{procaccia2006distortion}.

In the \emph{metric} distortion framework, introduced by Anshelevich, Bhardwaj, and Postl~\cite{anshelevich2015approximating,abp}, voters and candidates are points of an unknown metric space, the cost of a voter for a candidate is their distance, and every voter ranks the candidates from the closest to the farthest. The social cost of a candidate is its average distance to the voters. The metric assumption is motivated by spatial models of voting, and it makes the distortion of many natural voting rules bounded by a constant.

For deterministic rules, the picture is complete. Anshelevich et al.~\cite{abp} showed that no deterministic rule has distortion better than $3$, even when there are only two candidates. After a sequence of improvements~\cite{abp,munagala2019improved,kempe2020analytical}, Gkatzelis, Halpern, and Shah~\cite{ghs} designed a deterministic rule, called Plurality Matching, whose distortion is exactly $3$. Kizilkaya and Kempe~\cite{kk,kkgen} later obtained the same guarantee with a remarkably simple rule, called Plurality Veto.

Randomized rules can do better. Random Dictatorship, which elects the top choice of a uniformly random voter, has distortion $3-2/n$~\cite{anshelevich2017randomized}, and rules that only use plurality scores achieve $3-2/m$~\cite{kempe2020communication,ghs}; both bounds approach $3$ as the election grows. The simple example with two candidates and two voters shows that no randomized rule has distortion better than $2$~\cite{anshelevich2017randomized}, and Charikar and Ramakrishnan~\cite{cr} and Pulyassary and Swamy~\cite{pulyassary2021randomized} showed that $2$ itself is not achievable either; the lower bound of Charikar and Ramakrishnan is $2.1126$. On the positive side, Charikar, Ramakrishnan, Wang, and Wu~\cite{crww} gave the first randomized rule whose distortion is a constant below $3$, namely $2.753$. This bound was recently improved to $5/2$~\cite{frank,ye}, then to $2.1441$ by Shah~\cite{shah2441}, and then to $2.1248$ by Cai et al.~\cite{stable}.

The randomized rules mentioned above may put positive probability on many candidates, possibly including candidates that most voters find unacceptable. Such lotteries are hard to explain and to audit. Motivated by this, Cai, Gao, Ramakrishnan, and Wang~\cite{bounded} asked whether the barrier of $3$ can be broken with only \emph{bounded} randomness: are there absolute constants $\eps>0$ and $k$ such that some voting rule always outputs a uniform lottery over at most $k$ candidates and has distortion at most $3-\eps$? They answered this question affirmatively for some constant $k$, and conjectured that $k=2$ suffices. Jia and Peng~\cite{compression} later showed that a uniform list of $164$ candidates is enough.

\subsection{Our Contribution}
In this work, we show that the barrier of $3$ can be broken by randomizing over only \emph{two} candidates, provided that the two candidates may receive different probabilities. This is the smallest possible amount of randomness, since a single candidate cannot beat $3$.

\begin{theorem}[Main result, informal; proven in \cref{sec:overview,sec:computation}]\label{thm:informal}
There is an absolute constant $\eps_*>0$ such that for every election, there exist two candidates $a$ and $b$ and a probability $p\in[0,1]$ such that the lottery that elects $a$ with probability $p$ and $b$ with probability $1-p$ has metric distortion at most $3-\eps_*$. Moreover, such $a$, $b$, and $p$ can be computed in polynomial time.
\end{theorem}

\Cref{thm:informal} is a relaxation of the conjecture of Cai et al.~\cite{bounded}: the size of the support is two, as conjectured, but the two probabilities need not be equal. The uniform version of the conjecture remains open. The question of whether a lottery over two candidates can beat $3$ was also pointed out as open by Peters~\cite{peters}, who attributes it to Cai et al.~\cite{bounded} and Charikar et al.~\cite{crww}.

There is also a limit on what two candidates can achieve. Peters~\cite{peters} showed that the support of every lottery with distortion less than $3-\eps$ is $(1-\eps/2)$-undominated, that is, for every candidate $c$ outside the support, at most a $1-\eps/2$ fraction of the voters prefer $c$ to every member of the support. Combined with the classical fact that there are elections with no $\alpha$-undominated set of size two for any $\alpha<2/3$ (see the discussion in~\cite{bounded}), this shows that no two-candidate lottery can guarantee distortion better than $7/3$.

We give the formal statement of our result in \cref{sec:results}. The constant $\eps_*$ that our proof provides is extremely small, since the argument repeatedly uses widely separated scales; we made no attempt to optimize it.

We complement \cref{thm:informal} with a simple and explicit bound for elections in which some candidate is the top choice of many voters. If a candidate is ranked first by an $r$ fraction of the voters, then a two-candidate lottery achieves distortion $3-2r^2/(1-r+r^2)$ (\cref{thm:plurality}). For instance, this gives distortion $7/3$ when some candidate is the top choice of half of the voters. In contrast to the general case, this lottery is given by a short explicit formula.

\Cref{tab:results} summarizes the state of the art along with our result.

\begin{table}[h]
\centering
\renewcommand{\arraystretch}{1.25}
\begin{tabular}{|l|l|l|}
\hline
Reference & Type of rule & Distortion \TBstrut\\
\hline
\cite{abp,ghs,kk} & deterministic & $3$ (tight) \\
\cite{cr} & randomized, lower bound & $\geq 2.1126$\\
\cite{crww} & randomized & $2.753$\\
\cite{frank,ye} & randomized & $5/2$\\
\cite{shah2441} & randomized & $2.1441$\\
\cite{stable} & randomized & $2.1248$\\
\cite{bounded} & uniform over a constant number of candidates & $3-\Omega(1)$\\
\cite{compression} & uniform over $164$ candidates (with repetition) & $3-\Omega(1)$\\
\textbf{This work} & \textbf{two candidates, arbitrary probabilities} & $\mathbf{3-\Omega(1)}$\\
\hline
\end{tabular}
\caption{Metric distortion of single-winner voting rules. All bounds hold for general metric spaces.}\label{tab:results}
\end{table}

\subsection{Related Work}
The distortion of voting rules was introduced by Procaccia and Rosenschein~\cite{procaccia2006distortion}, and the metric version was introduced by Anshelevich et al.~\cite{anshelevich2015approximating,abp}. For deterministic rules, the conjecture that distortion $3$ is achievable was studied through several frameworks~\cite{munagala2019improved,kempe2020analytical} before it was resolved by Gkatzelis, Halpern, and Shah~\cite{ghs}. The rules of Kizilkaya and Kempe~\cite{kk,kkgen} are based on a continuous veto process, which is also one of the main tools of this paper. The analysis of all these rules is based on a matching certificate: a candidate has distortion at most $3$ if the voters can be matched to each other so that every voter weakly prefers this candidate to the top choice of its partner.

Among randomized rules, maximal lotteries~\cite{fishburn1984probabilistic} play a central role. They are equilibria of a symmetric zero-sum game in which the payoff of a candidate against another is the fraction of voters that prefer the first to the second. Charikar et al.~\cite{crww} showed that maximal lotteries have distortion $3$ and that a careful mixture of maximal lotteries with other rules beats $3$. Our pair selection procedure (\cref{sec:pair}) is based on a penalized variant of the same zero-sum game. Stable lotteries~\cite{stable} generalize maximal lotteries to lotteries over committees; in particular, Cai et al.~\cite{stable} showed that stable $2$-lotteries, which are distributions over pairs of candidates, have distortion $7/3$. We note that this is a different notion from ours: a stable $2$-lottery may be supported on many pairs, whereas our lottery is supported on two candidates.

A closely related line of work studies \emph{undominated committees}: small sets of candidates such that no outside candidate is preferred to all of them by a large fraction of the voters~\cite{six}. The existence proofs of such committees sample candidates from the equilibrium of a zero-sum game, and our \cref{lem:pair} is a penalized variant of this type of argument. Peters~\cite{peters} showed that the support of any lottery with distortion below $3$ is a constant-undominated committee, which connects the two lines of work.

The bounded randomness model was introduced by Cai et al.~\cite{bounded}, partly motivated by the use of uniform lotteries over panels in the selection of citizens' assemblies. Jia and Peng~\cite{compression} gave a general method for compressing a randomized rule into a uniform list of bounded length. Finally, metric distortion has also been studied for the selection of committees~\cite{multi,polar}, for elections in which voters may abstain~\cite{abstention}, for distributed elections in which groups of voters first select local representatives~\cite{distributed}, and for ballots that also report preference intensities~\cite{intensities}. We refer to the survey of Anshelevich et al.~\cite{survey} for a broader overview of distortion in social choice.

%% file: src/preliminaries.tex
\section{Preliminaries}\label{sec:prelim}

In this section, we define the metric distortion problem and fix the notation that we use throughout the paper.

\subsection{Elections and Metric Distortion}
An election consists of a finite set $V$ of $n$ voters and a finite set $C$ of $m$ candidates. Every voter $v\in V$ reports a strict ranking $\succ_v$ of the candidates. We write $c\succeq_v j$ if either $c=j$ or $c\succ_v j$. We denote the top choice of voter $v$ by $\tp(v)$. For a set $S\subseteq V$ of voters, we write $\mu(S)=|S|/n$ for the fraction of voters in $S$, and we write $\E_v$ for the expectation over a voter $v$ chosen uniformly at random from $V$. More generally, for a probability distribution $\mu'$ over $V$, we write $\mu'(S)=\sum_{v\in S}\mu'(v)$, and $\E_{v\sim\mu'}$ for the expectation over a voter $v$ drawn from $\mu'$. We write $\mathbf 1_E$ for the indicator of an event $E$. The \emph{plurality share} of a candidate $c$ is $\plu(c)=\mu(\{v:\tp(v)=c\})$, and for a set $I\subseteq C$ of candidates we write $\plu(I)=\sum_{c\in I}\plu(c)$.

A \emph{consistent metric} is a pseudometric $d$ on $V\cup C$ such that $d(v,c)\le d(v,j)$ for every voter $v$ and all candidates $c,j$ with $c\succ_v j$. In other words, voters rank the candidates by distance, and ties in distance are broken arbitrarily. Notice that if $d(v,c)<d(v,j)$, then $c\succ_v j$. The \emph{social cost} of a candidate $c$ is its average distance to the voters,
\[
\SC_d(c)=\E_v\, d(v,c),\qquad\text{and}\qquad \OPT_d=\min_{c\in C}\SC_d(c).
\]
A candidate $o$ with $\SC_d(o)=\OPT_d$ is called \emph{optimal}. We drop the subscript $d$ when the metric is clear from the context.

A \emph{lottery} $\cD$ is a probability distribution over the candidates, and its social cost is $\SC_d(\cD)=\E_{c\sim\cD}\SC_d(c)$. The \emph{distortion} of $\cD$ is
\[
\dist(\cD)=\sup_{d}\ \frac{\SC_d(\cD)}{\OPT_d},
\]
where the supremum is taken over all consistent metrics $d$ with $\OPT_d>0$. The distortion of a (randomized) voting rule is the supremum of $\dist(\cD)$ over all elections, where $\cD$ is the lottery that the rule outputs. A lottery with $\dist(\cD)\le\rho$ also satisfies $\SC_d(\cD)\le\rho\,\OPT_d$ for consistent metrics with $\OPT_d=0$. Indeed, increasing the distance between every two distinct points of $V\cup C$ by the same amount $\eta>0$ preserves both the triangle inequality and consistency, and makes the optimal cost positive; letting $\eta\to0$ gives the inequality.

We say that $\cD$ is a \emph{two-candidate lottery} if its support has at most two candidates. Such a lottery is described by two candidates $a,b$ and a probability $p$, and its social cost is $p\,\SC(a)+(1-p)\SC(b)$.

The social cost is invariant under scaling the metric, and so is the distortion. Therefore, whenever we fix a metric with $\OPT>0$ in the analysis, we scale it so that $\OPT=1$, and call it \emph{normalized}.

\subsection{Cuts and Couplings}
Our arguments repeatedly consider a set $I$ of candidates and ask how voters rank the members of $I$ relative to a fixed candidate. For a set $I\subseteq C$ and a candidate $c$, define
\[
\begin{aligned}
G_I&=\{v\in V:\tp(v)\in I\}, & s_I&=\mu(G_I)=\plu(I), & r_I&=1-s_I,\\
T_I(c)&=\{v\in V: j\succ_v c\text{ for every }j\in I\}, & t_I(c)&=\mu(T_I(c)).
\end{aligned}
\]
In words, $G_I$ is the set of voters whose top choice lies in $I$, and $T_I(c)$ is the set of voters who rank every member of $I$ above $c$. Notice that $T_I(c)=\emptyset$ whenever $c\in I$.

A \emph{coupling} is a nonnegative function $\pi:V\times V\to\R_{\ge0}$ such that $\sum_{u}\pi(v,u)=1/n$ for every $v$ and $\sum_{v}\pi(v,u)=1/n$ for every $u$. Equivalently, a coupling is a fractional perfect matching between the voters, scaled so that its total mass is one. We think of $\pi(v,u)$ as the mass of an edge from the \emph{source} $v$ to the \emph{destination} $u$. For sets $S,U\subseteq V$ we write $\pi(S,U)=\sum_{v\in S,u\in U}\pi(v,u)$. Notice that $\pi(S,U)\le\min\{\mu(S),\mu(U)\}$.

More generally, for probability distributions $\mu_1$ and $\mu_2$ over $V$, a \emph{coupling from $\mu_1$ to $\mu_2$} is a nonnegative function $\pi:V\times V\to\R_{\ge0}$ such that $\sum_{u}\pi(v,u)=\mu_1(v)$ for every $v$ and $\sum_{v}\pi(v,u)=\mu_2(u)$ for every $u$. Thus a coupling as defined above is a coupling from $\mu$ to $\mu$, where $\mu$ is the uniform distribution over $V$; unless stated otherwise, all couplings are of this form. For a coupling $\pi$ from $\mu_1$ to $\mu_2$, we have $\pi(S,U)\le\min\{\mu_1(S),\mu_2(U)\}$. Valid edges and the deficiency $\defi_c(\pi)$, defined below, are defined in the same way for such couplings.

An edge $(v,u)$ is \emph{valid} for a candidate $c$ if $c\succeq_v\tp(u)$, that is, if the source weakly prefers $c$ to the top choice of the destination. The \emph{deficiency} of a coupling $\pi$ for $c$ is the total mass of its invalid edges, and the deficiency of $c$ is the smallest deficiency over all couplings:
\[
\defi_c(\pi)=\sum_{(v,u):\ \tp(u)\succ_v c}\pi(v,u),\qquad \sigma(c)=\min_{\pi}\defi_c(\pi).
\]
The minimum exists since the set of couplings is compact. A candidate $c$ with $\sigma(c)=0$ is exactly a candidate that admits a perfect matching certificate in the sense of Gkatzelis, Halpern, and Shah~\cite{ghs}, and such a candidate has distortion at most $3$. We show in \cref{sec:profiles} that a candidate of small deficiency has distortion close to $3$ or smaller.

We think of a coupling $\pi$ with small deficiency for $c$ as a \emph{certificate} for $c$: all but a small mass of its edges $(v,u)$ are valid for $c$, that is, the source $v$ weakly prefers $c$ to the top choice of the destination $u$. A coupling with deficiency zero for $c$ is a perfect matching certificate for $c$.

Certificates are related to the sets defined at the beginning of this subsection as follows. Fix a set $I\subseteq C$ and a candidate $c$. We refer to the pair of sets $T_I(c)$ and $G_I$ as the \emph{cut} of $I$ for $c$, and to the quantity $r_I-t_I(c)$ as the \emph{slack} of this cut. A voter of $T_I(c)$ ranks every member of $I$ above $c$, so its edges to voters of $G_I$ are invalid for $c$. Hence, in a perfect matching certificate for $c$, the voters of $T_I(c)$ send all of their mass to voters outside $G_I$, whose total mass is $r_I$. Therefore, the slack of every cut is nonnegative for a candidate with deficiency zero. The following observation quantifies this for an arbitrary deficiency.

\begin{observation}\label{obs:cut}
For every coupling $\pi$, every candidate $c$, and every set $I\subseteq C$, we have $t_I(c)\le r_I+\defi_c(\pi)$.
\end{observation}
\begin{proof}
A voter $v\in T_I(c)$ ranks every member of $I$ above $c$, so every edge from $v$ to a voter of $G_I$ is invalid for $c$. Therefore
\begin{align*}
t_I(c)&=\pi\big(T_I(c),V\big)\\
&=\pi\big(T_I(c),G_I\big)+\pi\big(T_I(c),V\setminus G_I\big)\\
&\le\defi_c(\pi)+\pi\big(T_I(c),V\setminus G_I\big) &&\text{(edges from $T_I(c)$ to $G_I$ are invalid for $c$)}\\
&\le\defi_c(\pi)+\pi\big(V,V\setminus G_I\big)\\
&=\defi_c(\pi)+r_I.
\end{align*}
\end{proof}

We use $\log$ for the natural logarithm, and write $\pos{x}=\max\{x,0\}$.

%% file: src/results.tex
\section{Our Results and Techniques}\label{sec:results}

In this section, we state our results formally and explain the main ideas behind their proofs. Our main result is the following.

\begin{theorem}[proven in \cref{sec:overview,sec:computation}]\label{thm:main}
Define $\tau_{4000}=10^{-10}$, $\tau_j=\frac14\exp(-\tau_{j+1}^{-16})$ for $0\le j<4000$, and $\eps_*=\exp(-\tau_0^{-16})$. For every election, there is a two-candidate lottery $\cD$ such that
\[
\SC_d(\cD)\le(3-\eps_*)\,\OPT_d\qquad\text{for every consistent metric } d.
\]
Moreover, a two-candidate lottery of smallest distortion can be computed in time polynomial in $n$ and $m$.
\end{theorem}

We prove the existence part of \cref{thm:main} in \cref{sec:overview}, based on two lemmas that we prove in \cref{sec:descent}, and the algorithmic part in \cref{sec:computation}. Before explaining the ideas behind the proof, we present a much simpler result that already beats $3$ with two candidates in an important special case. It also illustrates how a two-candidate lottery can combine two complementary guarantees. We note that randomized rules that only use plurality scores were studied by Kempe~\cite{kempe2020communication} and Gkatzelis et al.~\cite{ghs}; their guarantees approach $3$ as the number of candidates grows, whereas the lottery below also uses the rankings through a veto step.

\subsection{Warm-up: Elections with a Popular Candidate}\label{sec:warmup}

We begin with a warm-up. \Cref{thm:plurality} below shows that if some candidate has a constant plurality share, then a lottery over two candidates achieves distortion bounded away from $3$. To the best of our knowledge, this has not appeared in previous work. We present it to illustrate the strength of the weighted veto lemma (\cref{lem:weighted}): combined with a few elementary applications of the triangle inequality, the lemma already yields a two-candidate lottery that beats distortion $3$ by a constant in an important special case. Extending this to all elections, however, requires substantially more structure, which we develop in the rest of the paper.

Suppose that some candidate $b$ is the top choice of an $r$ fraction of the voters. Candidate $b$ alone may have distortion close to $3$, and so may any other single candidate. However, the worst cases for $b$ and for a suitably chosen second candidate are different, and a lottery between the two balances them.

\begin{theorem}\label{thm:plurality}
If some candidate has plurality share $r>0$, then there is a two-candidate lottery with distortion at most
\[
3-\frac{2r^2}{1-r+r^2}.
\]
\end{theorem}
For example, the bound is $2.9780$ for $r=0.1$, $2.8462$ for $r=0.25$, and $7/3$ for $r=1/2$. The proof uses the following lemma, which is a weighted version of the certificate behind the Plurality Veto rule of Kizilkaya and Kempe~\cite{kk,kkgen}; we prove it in \cref{sec:certificates}.

\providecommand{\wvnote}{}\renewcommand{\wvnote}{; proven in \cref{sec:certificates}}\begin{restatable}[Weighted veto\wvnote]{lemma}{weightedveto}\label{lem:weighted}
Let $\mu_1$ and $\mu_2$ be two probability distributions over the voters. Then there exists a candidate $c$ with the following three properties.
\begin{enumerate}[label=(\alph*)]
\item $c$ is the top choice of some voter $u$ with $\mu_2(u)>0$.
\item There is a coupling from $\mu_1$ to $\mu_2$ all of whose edges are valid for $c$.
\item For every consistent metric $d$ and every candidate $o$,
\[
\E_{v\sim\mu_1}\big[d(v,c)-d(v,o)\big]\le 2\,\E_{u\sim\mu_2}\,d(u,o).
\]
\end{enumerate}
\end{restatable}

\begin{proof}[of \cref{thm:plurality}]
Let $b$ be a candidate with plurality share $r$ and let $U=G_{\{b\}}$ be the set of voters whose top choice is $b$. If $r=1$, then $b$ is the closest candidate of every voter and hence optimal, so assume $r<1$. Let $a$ be the candidate given by \cref{lem:weighted} when $\mu_1$ is uniform over $V$ and $\mu_2$ is uniform over $V\setminus U$.

Fix a consistent metric $d$ and an optimal candidate $o$, and define
\[
\alpha=\E_v\big[\mathbf 1_{v\in U}\, d(v,o)\big],\qquad \beta=\E_v\big[\mathbf 1_{v\notin U}\, d(v,o)\big],
\]
so that $\OPT=\alpha+\beta$. We first bound the cost of $a$:
\begin{align*}
\SC(a)-\OPT&=\E_{v\sim\mu_1}\big[d(v,a)-d(v,o)\big] &&\text{($\mu_1$ is uniform over $V$, and $\OPT=\SC(o)$)}\\
&\le 2\,\E_{u\sim\mu_2}\,d(u,o) &&\text{(\cref{lem:weighted}(c) with the metric $d$ and the candidate $o$)}\\
&=\frac{2}{(1-r)n}\sum_{u\in V\setminus U}d(u,o) &&\text{($\mu_2$ is uniform over $V\setminus U$, and $|V\setminus U|=(1-r)n$)}\\
&=\frac{2\beta}{1-r}.
\end{align*}
Thus $\SC(a)\le\OPT+2\beta/(1-r)$. To bound the cost of $b$, we first bound $d(b,o)$. Every voter $v\in U$ has $\tp(v)=b$, so consistency of $d$ gives $d(v,b)\le d(v,o)$. Hence, for every $v\in U$,
\begin{align*}
d(b,o)&\le d(b,v)+d(v,o)\\
&\le 2\,d(v,o) &&\text{(since $d(v,b)\le d(v,o)$)}.
\end{align*}
Averaging this inequality over $v\in U$, we obtain
\begin{align*}
d(b,o)&\le \frac{1}{|U|}\sum_{v\in U}2\,d(v,o)\\
&=\frac{2}{rn}\sum_{v\in U}d(v,o)\\
&=\frac{2\alpha}{r}.
\end{align*}
We now bound the cost of $b$:
\begin{align*}
\SC(b)&=\E_v\big[\mathbf 1_{v\in U}\,d(v,b)\big]+\E_v\big[\mathbf 1_{v\notin U}\,d(v,b)\big]\\
&\le \E_v\big[\mathbf 1_{v\in U}\,d(v,o)\big]+\E_v\big[\mathbf 1_{v\notin U}\,d(v,b)\big] &&\text{($d(v,b)\le d(v,o)$ for $v\in U$)}\\
&\le \E_v\big[\mathbf 1_{v\in U}\,d(v,o)\big]+\E_v\big[\mathbf 1_{v\notin U}\,\big(d(v,o)+d(o,b)\big)\big]\\
&=\E_v\,d(v,o)+\mu(V\setminus U)\, d(b,o)\\
&=\OPT+(1-r)\,d(b,o)\\
&\le\OPT+\frac{2(1-r)\alpha}{r} &&\text{(bound on $d(b,o)$ above)}.
\end{align*}
Now elect $b$ with probability $p_b=r/(1-r+r^2)$ and $a$ with probability $p_a=(1-r)^2/(1-r+r^2)$. These are probabilities, since $p_a+p_b=(1-2r+r^2+r)/(1-r+r^2)=1$. The expected excess cost over $\OPT$ satisfies
\begin{align*}
p_b\,\SC(b)+p_a\,\SC(a)-\OPT&=p_b\big(\SC(b)-\OPT\big)+p_a\big(\SC(a)-\OPT\big)\\
&\le p_b\cdot\frac{2(1-r)\alpha}{r}+p_a\cdot\frac{2\beta}{1-r}\\
&=\frac{2(1-r)\,\alpha}{1-r+r^2}+\frac{2(1-r)\,\beta}{1-r+r^2}\\
&=\frac{2(1-r)}{1-r+r^2}\,\OPT.
\end{align*}
Dividing by $\OPT>0$, the distortion is at most
\begin{align*}
1+\frac{2(1-r)}{1-r+r^2}&=\frac{3-3r+r^2}{1-r+r^2}\\
&=3-\frac{2r^2}{1-r+r^2}.
\end{align*}
\end{proof}

\Cref{thm:plurality} gives a constant improvement only when some candidate has a constant plurality share. The following example shows that in general, one cannot restrict attention to candidates that are the top choice of some voter.

\begin{example}\label{ex:star}
Consider $n$ voters $v_1,\ldots,v_n$ and candidates $z,a_1,\ldots,a_n$. Voter $v_i$ ranks $a_i$ first, $z$ second, and the remaining candidates afterwards. Place $z$ at the center of a star with $n$ arms of length $1$, put $a_i$ at the end of the $i$-th arm, and put $v_i$ at the middle of the $i$-th arm. This metric is consistent (voter $v_i$ is at distance $1/2$ from both $a_i$ and $z$). We have $\SC(z)=1/2$, since every voter is at distance $1/2$ from $z$. For every $i$,
\begin{align*}
\SC(a_i)&=\frac1n\Big(d(v_i,a_i)+\sum_{k\neq i}d(v_k,a_i)\Big)\\
&=\frac1n\Big(\frac12+(n-1)\Big(\frac12+1\Big)\Big) &&\text{($v_k$ reaches $a_i$ through $z$ for $k\neq i$)}\\
&=\frac32-\frac1n.
\end{align*}
The candidates with positive plurality share are exactly $a_1,\ldots,a_n$. Hence every lottery $\cD$ supported on them satisfies
\begin{align*}
\frac{\SC(\cD)}{\OPT}&\ge\frac{\SC(\cD)}{\SC(z)}\\
&=\frac{3/2-1/n}{1/2}\\
&=3-\frac2n.
\end{align*}
Therefore, every lottery supported on candidates with positive plurality share has distortion at least $3-2/n$.
\end{example}

\subsection{Overview of the Proof of \Cref{thm:main}}\label{sec:overview}

The warm-up relies on a single popular candidate $b$: all voters of $U=G_{\{b\}}$ rank $b$ first. In a general election, no candidate needs to have a large plurality share. Our proof replaces the popular candidate by a popular \emph{group} of candidates, namely a set $I_0$ whose supporters rank it above all candidates that are far from it. Such a set need not be exactly closed, so we allow a buffer of candidates with a small plurality share. This leads to the following definition, which depends only on the rankings.

\begin{definition}[Fuzzy block]\label{def:block}
Let $0<\tau\le10^{-10}$. A \emph{$\tau$-block} is a triple of sets of candidates $I_0\subseteq I_1\subseteq I_2$ with the following properties. Let $G=G_{I_0}$ be the set of voters whose top choice lies in $I_0$. Then $\plu(I_2\setminus I_0)\le2\tau^2$, and all voters of $G$ except for a set of mass at most $\tau^2$ satisfy
\begin{equation}\label{eq:blockorder}
I_0\succ_v C\setminus I_1\qquad\text{and}\qquad I_1\succ_v C\setminus I_2,
\end{equation}
where, for sets $S,S'\subseteq C$, $S\succ_v S'$ means that $v$ ranks every member of $S$ above every member of $S'$. We call the voters of $G$ that satisfy \eqref{eq:blockorder} the \emph{core voters} of the block, and $s_{I_0}=\mu(G)$ the \emph{mass} of the block.
\end{definition}

Notice that a block is a purely ordinal object: it depends only on the rankings. In particular, a block extracted from one metric remains a block when we later consider a different metric.

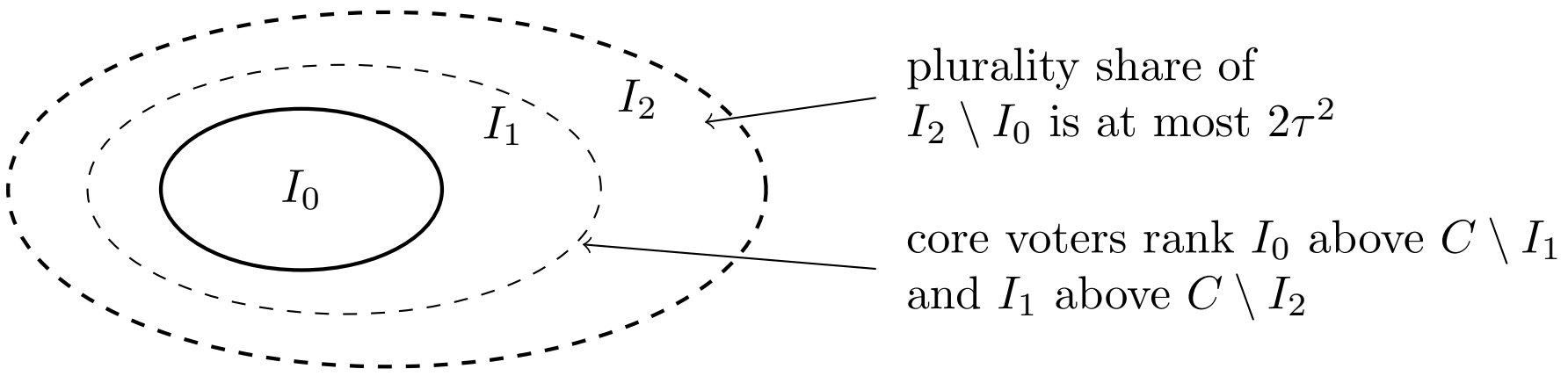
\begin{figure}[h]
\centering
\begin{tikzpicture}[font=\small,scale=.92]
\draw[dashed,thick] (0,0) ellipse (3.1 and 1.45);
\draw[dashed] (-.35,0) ellipse (2.1 and 1.02);
\draw[thick] (-.7,0) ellipse (1.15 and .66);
\node at (-.7,0) {$I_0$};\node at (.95,.52) {$I_1$};\node at (2.05,.74) {$I_2$};
\node[align=left,anchor=west] at (4.1,.75) {plurality share of\\$I_2\setminus I_0$ is at most $2\tau^2$};
\node[align=left,anchor=west] at (4.1,-.65) {core voters rank $I_0$ above $C\setminus I_1$\\and $I_1$ above $C\setminus I_2$};
\draw[->] (4.0,.75)--(2.6,.55);\draw[->] (4.0,-.65)--(1.6,-.45);
\end{tikzpicture}
\caption{A fuzzy block. The two outer layers absorb a small plurality share, so no set needs to be exactly closed.}\label{fig:block}
\end{figure}

The proof of \cref{thm:main} is based on the following two lemmas. Fix an election, and recall the parameters $\tau_0,\ldots,\tau_{4000}$ and $\eps_*$ of \cref{thm:main}.

\providecommand{\abnote}{}\renewcommand{\abnote}{; proven in \cref{sec:descent}}
\begin{restatable}[Initial block\abnote]{lemma}{initialblock}\label{lem:initial}
Either some two-candidate lottery has distortion at most $3-\eps_*$, or there is a $\tau_0$-block with mass in $[1/20,3/4]$.
\end{restatable}

\begin{restatable}[Descent step\abnote]{lemma}{descentstep}\label{lem:step}
Let $0\le j<4000$. Either some two-candidate lottery has distortion at most $3-\eps_*$, or for every $\tau_j$-block with mass $s\in[1/20,3/4]$, there is a $\tau_{j+1}$-block with mass $s'\in[1/20,3/4]$ and $s'\le s-1/4000$.
\end{restatable}

\begin{proof}[of the existence part of \cref{thm:main}]
Suppose that no two-candidate lottery has distortion at most $3-\eps_*$. By \cref{lem:initial}, there is a $\tau_0$-block with mass $s_0\in[1/20,3/4]$. Applying \cref{lem:step} for $j=0,1,\ldots,2999$, we obtain $\tau_j$-blocks with masses $s_j\in[1/20,3/4]$ such that $s_j\le s_{j-1}-1/4000$ for every $j\ge1$. Hence
\[
\frac1{20}\le s_{3000}\le s_0-\frac{3000}{4000}\le\frac34-\frac34=0,
\]
a contradiction. Therefore some two-candidate lottery has distortion at most $3-\eps_*$.
\end{proof}

We prove \cref{lem:initial,lem:step} in \cref{sec:descent}, from four building blocks that we state in \cref{sec:building} and prove in \cref{sec:certificates,sec:repair,sec:pairtoblock,sec:coupled}.

%% file: src/building.tex
\subsection{The Building Blocks}\label{sec:building}

We prove \cref{lem:initial,lem:step} from the weighted veto lemma (\cref{lem:weighted}) and three further lemmas, which we state now. Each of them is proven in its own section, together with the definitions that its proof requires. Given these four lemmas, the proofs of \cref{lem:initial,lem:step} only count coupling mass along the core voters of blocks; in particular, they do not involve metrics.

The first lemma replaces a candidate $a$ and its certificate by a candidate $c$ inside a given set $I$ and a certificate for $c$, at the cost of the slack $r_I-t_I(a)$ of the cut of $I$ for $a$.

\gdef\repnote{; proven in \cref{sec:repair}}
\begin{restatable}[Repair\repnote]{lemma}{repairlemma}\label{lem:repair}
Let $\pi$ be a coupling, let $a$ be a candidate, and let $I\subseteq C$ be nonempty. There is a candidate $c\in I$ and a coupling $\pi'$ such that
\[
\defi_c(\pi')\le\defi_a(\pi)+r_I-t_I(a).
\]
Moreover, for every set $G\subseteq V$, we have $\pi'(G,G)\ge\pi(G,G)-\mu(G_I\setminus G)$.
\end{restatable}

The second lemma is our source of blocks. Given an upper bound $\delta$ on the deficiencies of the candidates, it selects a two-candidate lottery such that, if this lottery has distortion close to $3$, then there is a block $(J_0,J_1,J_2)$ such that the slack of the cut of $J_0$ is tiny for both candidates of the lottery.

\gdef\ptbnote{; proven in \cref{sec:pairtoblock}}
\begin{restatable}[Pair-to-block\ptbnote]{lemma}{pairtoblocklemma}\label{lem:pairtoblock}
Let $0<\tau\le10^{-10}$, let $\delta:C\to\R_{\ge0}$ satisfy $\delta(c)\ge\sigma(c)$ for every candidate $c$, and let $\xi$ and $\eps$ satisfy $0<\eps\le\xi\le\exp(-\tau^{-16})$ and $\min_c\delta(c)\le\xi$. There are candidates $a,b$ and a probability $p\in[0,1]$ with the following property. If the lottery that elects $a$ with probability $p$ and $b$ with probability $1-p$ has distortion larger than $3-\eps$, then there exist a $\tau$-block $(J_0,J_1,J_2)$ and a number $\zeta\in(0,1)$ such that:
\begin{enumerate}
\item $a,b\notin J_2$;
\item for both $i\in\{a,b\}$, we have $-\delta(i)\le r_{J_0}-t_{J_0}(i)\le48\,(\delta(i)+\xi)/\zeta\le\tau^2$ and $\delta(i)+\xi\le\tau^2$;
\item the mass of the block satisfies $s_{J_0}\le3/4$;
\item if $a'\in\{a,b\}$ satisfies $\delta(a')=\min\{\delta(a),\delta(b)\}$, and some candidate $c\in J_0$ satisfies $\delta(c)+\xi\le50\,(\delta(a')+\xi)/\zeta$, then $s_{J_0}\ge1/20$.
\end{enumerate}
\end{restatable}

The third lemma shows that a block yields a candidate with a certificate of small deficiency that keeps a constant amount of mass inside the voters of the block.

\gdef\pmnote{; proven in \cref{sec:coupled}}
\begin{restatable}[Protected matching\pmnote]{lemma}{protectedlemma}\label{lem:protected}
Let $0<\tau\le10^{-10}$, let $(I_0,I_1,I_2)$ be a $\tau$-block with mass $s\in[1/20,3/4]$, let $G=G_{I_0}$, and let $0<\eps\le\exp(-\tau^{-16})$. Either some two-candidate lottery has distortion at most $3-\eps$, or there is a candidate $c$ and a coupling $\pi'$ such that
\[
\defi_c(\pi')\le2\tau^2\qquad\text{and}\qquad\pi'(G,G)\ge s^2/4.
\]
\end{restatable}

In \cref{sec:descent}, we prove \cref{lem:initial,lem:step} from these four lemmas. We briefly summarize the argument. Suppose that every two-candidate lottery has distortion larger than $3-\eps_*$.

For \cref{lem:initial}, we apply \cref{lem:pairtoblock} with $\delta=\sigma$; its hypothesis on $\min_c\delta(c)$ holds since, by \cref{lem:weighted}, some candidate has deficiency zero. This gives a block of mass at most $3/4$. Using \cref{lem:repair}, we replace the candidate with the smaller deficiency by a candidate inside the block at a small cost, and property (iv) of \cref{lem:pairtoblock} shows that the mass of the block is at least $1/20$.

For \cref{lem:step}, fix a $\tau_j$-block, and let $G$ be the set of voters whose top choice lies in its innermost set. We measure the quality of a candidate by a deficiency that also charges for a loss of coupling mass inside $G$. By \cref{lem:protected}, some candidate is good in this measure. Applying \cref{lem:pairtoblock} to this measure gives two candidates and a new $\tau_{j+1}$-block. Counting coupling mass along the core voters of the two blocks shows that both candidates lie in the old block, and that the new block lies inside the old one up to its buffer. The certificate of the better candidate keeps a constant amount of mass inside $G$, but almost none of it inside the voters of the new block, so the new block misses a constant part of $G$ and its mass decreases by at least $1/4000$. Finally, as in \cref{lem:initial}, \cref{lem:repair} and property (iv) show that the mass of the new block is still at least $1/20$.

%% file: src/veto.tex
\section{Proof of the Weighted Veto Lemma}\label{sec:certificates}

In this section, we prove the weighted veto lemma (\cref{lem:weighted}), stated in \cref{sec:warmup}. For $\mu_1=\mu_2$ uniform, the procedure in the proof is the Plurality Veto rule of Kizilkaya and Kempe~\cite{kk,kkgen} in continuous time.

\renewcommand{\wvnote}{}\weightedveto*
\begin{proof}
Give every candidate $j$ an initial score $h_0(j)=\mu_2(\{u:\tp(u)=j\})$. The scores sum to one. At every time $t\in[0,1)$, let $R_t$ be the set of candidates with positive score. Every voter $v$ points to its least preferred candidate in $R_t$ and decreases the score of that candidate at rate $\mu_1(v)$. The total score decreases at rate one, and therefore all scores reach zero exactly at time one. The process is well-defined, since between two consecutive times at which a score reaches zero, every voter points to a fixed candidate and all scores decrease linearly. Let $c$ be a candidate whose score reaches zero last (ties broken arbitrarily); we show that $c$ satisfies the claims of the lemma. Then $c\in R_t$ for all $t<1$, and $c$ is the top choice of some voter in the support of $\mu_2$.

We build the coupling as follows. Whenever voter $v$ decreases the score of a candidate $j$ by some amount, we add this amount of mass to edges from $v$ to the voters $u$ with $\tp(u)=j$, in proportion to $\mu_2(u)$. Every voter $v$ sends a total mass of $\mu_1(v)$, and since the score of every candidate is eventually exhausted, every voter $u$ receives a total mass of $\mu_2(u)$. Thus we obtain a coupling from $\mu_1$ to $\mu_2$. If $v$ sends mass to $u$ at time $t$, then $\tp(u)$ is the least preferred candidate of $v$ in $R_t$, and $c\in R_t$. Hence $c\succeq_v\tp(u)$, that is, the edge is valid for $c$.

To prove (c), consider an edge $(v,u)$ of the coupling constructed above; by (b), this edge is valid for $c$. Fix a consistent metric $d$ and a candidate $o$. Then
\begin{align*}
d(v,c)&\le d\big(v,\tp(u)\big) &&\text{(consistency, since $c\succeq_v\tp(u)$)}\\
&\le d(v,o)+d(o,u)+d\big(u,\tp(u)\big)\\
&\le d(v,o)+d(o,u)+d(u,o) &&\text{(consistency, since $\tp(u)\succeq_u o$)}\\
&=d(v,o)+2\,d(u,o).
\end{align*}
Let $\pi$ denote the coupling constructed above. Then
\begin{align*}
\E_{v\sim\mu_1}\big[d(v,c)-d(v,o)\big]&=\sum_{v,u}\pi(v,u)\big(d(v,c)-d(v,o)\big)\\
&\le\sum_{v,u}\pi(v,u)\cdot2\,d(u,o)\\
&=2\,\E_{u\sim\mu_2}\,d(u,o).
\end{align*}
\end{proof}

In particular, for $\mu_1=\mu_2=\mu$ uniform, some candidate has deficiency zero.

%% file: src/repair.tex
\section{Proof of the Repair Lemma}\label{sec:repair}

In this section, we prove the repair lemma (\cref{lem:repair}). We prove a more general version, which is also used in \cref{sec:coupled}, in which the edges of a coupling may carry different candidate labels. A \emph{labeled coupling} is a finite family of nonnegative functions $\pi_1,\ldots,\pi_k$ on $V\times V$ with labels $c_1,\ldots,c_k\in C$, such that $\pi=\pi_1+\cdots+\pi_k$ is a coupling. Its \emph{invalid mass} is $\sum_{i}\defi_{c_i}(\pi_i)$, and for a set $I\subseteq C$ its \emph{cut mass} is
\[
\tbar_\pi(I)=\sum_{i=1}^k\pi_i\big(T_I(c_i),V\big),
\]
the mass of edges whose source ranks all of $I$ above the label of the edge. An ordinary coupling $\pi$ for a candidate $a$ is a labeled coupling with $k=1$, and then its invalid mass is $\defi_a(\pi)$ and its cut mass is $t_I(a)$.

\begin{lemma}[Repair, labeled version]\label{lem:repairlabeled}
Let $\pi_1,\ldots,\pi_k$ be a labeled coupling with invalid mass $\sigma_0$ and let $I\subseteq C$ be nonempty. There is a candidate $c\in I$ and a coupling $\pi'$ such that
\[
\defi_c(\pi')\le \sigma_0+r_I-\tbar_\pi(I).
\]
Moreover, for every set $G\subseteq V$ we have $\pi'(G,G)\ge \pi(G,G)-\mu(G_I\setminus G)$, where $\pi=\pi_1+\cdots+\pi_k$.
\end{lemma}
\begin{proof}
If $G_I=\emptyset$, we let $c$ be an arbitrary member of $I$ and $\pi'=\pi$; the counting argument below applies verbatim (with no removed edges). So assume $G_I\neq\emptyset$.

Remove all edges of $\pi$ whose destination lies in $G_I$, and let $\mu_1'$ be the distribution of the sources of the removed mass, normalized to a probability distribution, and $\mu_2'$ the uniform distribution over $G_I$ (the removed mass enters every voter of $G_I$ completely). Apply \cref{lem:weighted} with $\mu_1=\mu_1'$ and $\mu_2=\mu_2'$. It gives a candidate $c$ which is the top choice of a voter of $G_I$, so $c\in I$, and a coupling from $\mu_1'$ to $\mu_2'$ all of whose edges are valid for $c$. Scale this coupling by $\mu(G_I)$ and add it back to the remaining edges, to obtain $\pi'$. All new edges are valid for $c$, and every source sends exactly the same mass to $G_I$ as before, while every voter of $G_I$ receives a mass of $\mu(G_I)/|G_I|=1/n$. Hence $\pi'$ is a coupling.

It remains to count the invalid edges among the retained edges, which all go to $V\setminus G_I$. Consider a retained edge with label $c_i$ whose source $v$ lies in $T_I(c_i)$. Then $v$ ranks $c\in I$ above $c_i$, so if the edge was valid for $c_i$, it is also valid for $c$. Hence the retained edges that are invalid for $c$ are either invalid for their label, or have a source outside $T_I(\text{label})$. Define
\[
\chi^-=\sum_{i}\pi_i\big(T_I(c_i),G_I\big),\qquad \chi^+=\sum_i\pi_i\big(V\setminus T_I(c_i),V\setminus G_I\big).
\]
The edges counted in $\chi^-$ are invalid for their labels, since their destinations have top choices in $I$. They are removed, so the retained edges that are invalid for their label have mass at most $\sigma_0-\chi^-$. Therefore $\defi_c(\pi')\le \sigma_0-\chi^-+\chi^+$. Write $\chi^0=\sum_i\pi_i\big(T_I(c_i),V\setminus G_I\big)$. Then
\begin{align*}
r_I&=\pi\big(V,V\setminus G_I\big)\\
&=\chi^0+\chi^+,\\
\tbar_\pi(I)&=\sum_i\pi_i\big(T_I(c_i),G_I\big)+\chi^0\\
&=\chi^-+\chi^0.
\end{align*}
Subtracting the two identities gives $\chi^+-\chi^-=r_I-\tbar_\pi(I)$. Hence
\begin{align*}
\defi_c(\pi')&\le\sigma_0-\chi^-+\chi^+\\
&=\sigma_0+r_I-\tbar_\pi(I),
\end{align*}
which proves the first claim.

For the second claim, notice that $\pi'(G,G_I)=\pi(G,G_I)$ since every source keeps its mass to $G_I$, and $\pi'(G,G_I\setminus G)\le\mu(G_I\setminus G)$ by the capacity of the destinations. Also, the edges into $G\setminus G_I$ are unchanged. Hence
\begin{align*}
\pi'(G,G)&=\pi'(G,G\setminus G_I)+\pi'(G,G_I)-\pi'(G,G_I\setminus G)\\
&=\pi(G,G\setminus G_I)+\pi(G,G_I)-\pi'(G,G_I\setminus G)\\
&\ge\pi(G,G\setminus G_I)+\pi(G,G_I)-\mu(G_I\setminus G) &&\text{(capacity of the destinations)}\\
&\ge\pi(G,G\setminus G_I)+\pi(G,G\cap G_I)-\mu(G_I\setminus G)\\
&=\pi(G,G)-\mu(G_I\setminus G).\qedhere
\end{align*}
\end{proof}

We now restate and prove \cref{lem:repair}.

\gdef\repnote{}\repairlemma*
\begin{proof}
This is the special case $k=1$ of \cref{lem:repairlabeled}, with the single label $a$: the coupling $\pi$ has invalid mass $\defi_a(\pi)$ and cut mass $t_I(a)$.
\end{proof}

%% file: src/profiles.tex
\section{Proof of the Pair-to-Block Lemma}\label{sec:pairtoblock}

In this section, we prove \cref{lem:pairtoblock}. In \cref{sec:profiles}, we describe a fixed metric through distances to an optimal candidate, and relate the cost of a candidate to its cuts. In \cref{sec:pair}, we define the penalized pair, which gives the candidates $a$ and $b$ of the lemma. In \cref{sec:blocks}, we extract a fuzzy block from a bad metric for the penalized pair, and in \cref{sec:p2bproof} we complete the proof.

\subsection{Bad Metrics and Distance Profiles}\label{sec:profiles}

Let $\eps>0$. A \emph{bad metric} for a two-candidate lottery $\cD$ is a normalized consistent metric $d$ with $\SC_d(\cD)\ge3-\eps$. If $\cD$ has distortion larger than $3-\eps$, then there is a consistent metric $d$ with $\OPT_d>0$ and $\SC_d(\cD)>(3-\eps)\OPT_d$; after scaling, $d$ is a bad metric for $\cD$.

Whenever we fix a normalized consistent metric and an optimal candidate $o$ (so that $\SC(o)=1$), here and in \cref{sec:coupled}, we use the following notation. For every candidate $c$, every voter $v$, and every $\ell\ge0$, define
\[
x_c=d(c,o),\qquad y_v=d(v,o),\qquad z_v=x_{\tp(v)},\qquad e_v(c)=\min_{j:\,c\succeq_v j}x_j,\qquad I_\ell=\{c\in C:x_c\le \ell\}.
\]
Thus $\E_v y_v=1$, the quantity $z_v$ is the distance from the top choice of $v$ to $o$, and $e_v(c)$ is the distance to $o$ of the closest candidate that $v$ ranks weakly below $c$. Notice that $o\in I_\ell$ for all $\ell\ge0$.

\begin{lemma}\label{lem:profile}
For every candidate $c$ and every voter $v$,
\[
d(v,c)-y_v\le e_v(c)\le x_c,\qquad z_v\le 2y_v,\qquad \E_v z_v\le 2.
\]
Moreover, for every two candidates $a,b$,
\[
\E_v e_v(c)=\int_0^\infty t_{I_\ell}(c)\,d\ell,\qquad \E_v z_v=\int_0^\infty r_{I_\ell}\,d\ell,\qquad \E_v|e_v(a)-e_v(b)|=\int_0^\infty\mu\big(T_{I_\ell}(a)\,\triangle\, T_{I_\ell}(b)\big)\,d\ell.
\]
In particular, $\SC(c)\le 1+\E_v e_v(c)$.
\end{lemma}
\begin{proof}
Let $j$ be a candidate with $c\succeq_v j$. Then
\begin{align*}
d(v,c)&\le d(v,j) &&\text{(consistency, since $c\succeq_v j$)}\\
&\le d(v,o)+d(o,j)\\
&=y_v+x_j.
\end{align*}
Minimizing over $j$ gives $d(v,c)-y_v\le e_v(c)$, and choosing $j=c$ gives $e_v(c)\le x_c$. Next,
\begin{align*}
z_v&=d\big(\tp(v),o\big)\\
&\le d\big(\tp(v),v\big)+d(v,o)\\
&\le d(v,o)+d(v,o) &&\text{(consistency, since $\tp(v)\succeq_v o$)}\\
&=2y_v.
\end{align*}
Averaging over $v$ and using $\E_v y_v=\SC(o)=1$ gives $\E_v z_v\le2$.

For the identities, observe that $e_v(c)>\ell$ if and only if no candidate in $I_\ell$ is ranked weakly below $c$ by $v$, that is, if and only if $v\in T_{I_\ell}(c)$. Similarly, $z_v>\ell$ if and only if $v\notin G_{I_\ell}$. Every nonnegative number $w$ satisfies $w=\int_0^\infty\mathbf 1_{w>\ell}\,d\ell$. Hence
\begin{align*}
\E_v e_v(c)&=\E_v\int_0^\infty\mathbf 1_{e_v(c)>\ell}\,d\ell\\
&=\int_0^\infty\mu\big(\{v:e_v(c)>\ell\}\big)\,d\ell\\
&=\int_0^\infty t_{I_\ell}(c)\,d\ell &&\text{(since $e_v(c)>\ell$ if and only if $v\in T_{I_\ell}(c)$)}.
\end{align*}
In the same way,
\begin{align*}
\E_v z_v&=\int_0^\infty\mu\big(\{v:z_v>\ell\}\big)\,d\ell\\
&=\int_0^\infty r_{I_\ell}\,d\ell &&\text{(since $z_v>\ell$ if and only if $v\notin G_{I_\ell}$)}.
\end{align*}
For the third identity, every two nonnegative numbers $w_1,w_2$ satisfy $|w_1-w_2|=\int_0^\infty|\mathbf 1_{w_1>\ell}-\mathbf 1_{w_2>\ell}|\,d\ell$. Hence
\begin{align*}
\E_v|e_v(a)-e_v(b)|&=\int_0^\infty\E_v\big|\mathbf 1_{e_v(a)>\ell}-\mathbf 1_{e_v(b)>\ell}\big|\,d\ell\\
&=\int_0^\infty\mu\big(T_{I_\ell}(a)\,\triangle\,T_{I_\ell}(b)\big)\,d\ell &&\text{(by the two equivalences above)}.
\end{align*}
Finally,
\begin{align*}
\SC(c)&=\E_v d(v,c)\\
&\le\E_v\big(y_v+e_v(c)\big) &&\text{(first inequality of the lemma)}\\
&=1+\E_v e_v(c).
\end{align*}
\end{proof}

Combining \cref{obs:cut} with \cref{lem:profile} shows that a candidate with deficiency zero has cost at most $1+\int_0^\infty r_{I_\ell}\,d\ell\le 3$. The next lemma quantifies this for candidates with small deficiency, and shows that a candidate whose cost is close to $3$ makes almost all threshold cuts nearly tight. For a candidate $c$, we write
\[
g_c(\ell)=r_{I_\ell}-t_{I_\ell}(c)\qquad(\ell\ge0)
\]
for the slack of the cut inequality at threshold $\ell$.

\begin{lemma}[Near-tight candidates]\label{lem:cost}
Let $c$ be a candidate with $\sigma(c)<1$. Then
\[
x_c\le\frac{4}{1-\sigma(c)},\qquad \SC(c)\le 3+\sigma(c)\,x_c\le 3+\frac{4\sigma(c)}{1-\sigma(c)}.
\]
If in addition $\SC(c)\ge 3-\alpha$ for some $\alpha\ge0$, then
\[
\int_0^\infty\pos{g_c(\ell)}\,d\ell\le\alpha+\sigma(c)\,x_c\qquad\text{and}\qquad \E_v(2y_v-z_v)\le\alpha+\sigma(c)\,x_c.
\]
\end{lemma}
\begin{proof}
Let $\pi$ be a coupling with $\defi_c(\pi)=\sigma(c)$. On a valid edge $(v,u)$, we have $c\succeq_v\tp(u)$, and therefore $e_v(c)\le x_{\tp(u)}=z_u$. On an invalid edge, we use $e_v(c)\le x_c$. Hence
\begin{align*}
\E_v e_v(c)&=\sum_{v,u}\pi(v,u)\,e_v(c)\\
&\le\sum_{(v,u)\text{ valid}}\pi(v,u)\,z_u+\sum_{(v,u)\text{ invalid}}\pi(v,u)\,x_c &&\text{(bounds on $e_v(c)$ above)}\\
&\le\sum_{v,u}\pi(v,u)\,z_u+\sigma(c)\,x_c &&\text{(since $z_u\ge0$ and $\defi_c(\pi)=\sigma(c)$)}\\
&=\E_u z_u+\sigma(c)\,x_c\\
&\le2+\sigma(c)\,x_c &&\text{(\cref{lem:profile})}.
\end{align*}
Consequently,
\begin{align*}
\SC(c)&\le1+\E_v e_v(c) &&\text{(\cref{lem:profile})}\\
&\le3+\sigma(c)\,x_c.
\end{align*}
For the bound on $x_c$,
\begin{align*}
x_c&=\E_v\,d(c,o)\\
&\le\E_v\big(d(c,v)+d(v,o)\big)\\
&=\SC(c)+1\\
&\le4+\sigma(c)\,x_c &&\text{(bound on $\SC(c)$ above)}.
\end{align*}
Since $\sigma(c)<1$, rearranging gives $x_c\le4/(1-\sigma(c))$, and substituting this bound into $\SC(c)\le3+\sigma(c)x_c$ gives $\SC(c)\le3+4\sigma(c)/(1-\sigma(c))$.

Now suppose that $\SC(c)\ge3-\alpha$. By \cref{obs:cut}, $g_c(\ell)\ge-\sigma(c)$ for every $\ell$. Moreover, for $\ell\ge x_c$ we have $c\in I_\ell$, so $t_{I_\ell}(c)=0$ and $g_c(\ell)\ge0$. Writing $\pos{-g_c(\ell)}$ for the negative part of $g_c(\ell)$, we obtain $\int_0^\infty\pos{-g_c(\ell)}\,d\ell\le\sigma(c)\,x_c$. Moreover,
\begin{align*}
\int_0^\infty g_c(\ell)\,d\ell&=\int_0^\infty r_{I_\ell}\,d\ell-\int_0^\infty t_{I_\ell}(c)\,d\ell\\
&=\E_v z_v-\E_v e_v(c) &&\text{(\cref{lem:profile})}\\
&\le2-\big(\SC(c)-1\big) &&\text{(\cref{lem:profile})}\\
&\le\alpha &&\text{(since $\SC(c)\ge3-\alpha$)}.
\end{align*}
Therefore,
\begin{align*}
\int_0^\infty\pos{g_c(\ell)}\,d\ell&=\int_0^\infty g_c(\ell)\,d\ell+\int_0^\infty\pos{-g_c(\ell)}\,d\ell\\
&\le\alpha+\sigma(c)\,x_c.
\end{align*}
For the second inequality,
\begin{align*}
\E_v(2y_v-z_v)&=2-\E_v z_v\\
&\le2-\E_v e_v(c)+\sigma(c)\,x_c &&\text{(the first chain of inequalities in this proof)}\\
&\le2-\big(\SC(c)-1\big)+\sigma(c)\,x_c &&\text{(\cref{lem:profile})}\\
&\le\alpha+\sigma(c)\,x_c &&\text{(since $\SC(c)\ge3-\alpha$)}.
\end{align*}
\end{proof}

%% file: src/pair.tex
\subsection{Penalized Pair Selection}\label{sec:pair}

In this subsection, we describe how we choose the two candidates of our lotteries. The choice is based on a symmetric zero-sum game in which candidates are compared by majority, as in maximal lotteries~\cite{fishburn1984probabilistic}, and every candidate additionally pays a penalty. We first prove a general statement for an arbitrary penalty, and then specialize it to logarithmic penalties of deficiencies.

\begin{lemma}[Penalized pair selection]\label{lem:pair}
Let $\phi:C\to\R_{\ge0}$ be a penalty function and let $\gamma>0$. There are candidates $a,b$ (possibly equal) such that, writing $\bar\phi=(\phi(a)+\phi(b))/2$,
\begin{align}
\bar\phi&\le\min_{c\in C}\phi(c)+\frac{5}{6\gamma},\label{eq:paircost}\\
1-q(c)+\gamma\big(\phi(c)-\bar\phi\big)&\ge\frac16\qquad\text{for every candidate }c,\label{eq:pairdom}
\end{align}
where $q(c)=\mu(\{v: c\succ_v a\text{ and }c\succ_v b\})$ is the fraction of voters that prefer $c$ to both $a$ and $b$.
\end{lemma}

For $\phi\equiv0$, the lemma says that two independent samples from a maximal lottery form a pair such that no candidate is preferred to both by more than a $5/6$ fraction of the voters; arguments of this type are standard in the study of undominated committees~\cite{six}. In words, the average penalty of the pair is nearly the smallest possible, and a candidate that is preferred to both $a$ and $b$ by almost all voters must have a penalty noticeably larger than that of the pair.

\begin{proof}[of \cref{lem:pair}]
Consider the zero-sum game in which both players choose a candidate, and the payoff of the first player when the choices are $c$ and $j$ is
\[
\psi(c,j)=\mu(\{v:c\succ_v j\})+\tfrac12\mathbf 1_{c=j}-\tfrac12-\gamma\big(\phi(c)-\phi(j)\big).
\]
Since $\mu(\{v:c\succ_v j\})+\mu(\{v:j\succ_v c\})=1$ for $c\neq j$, we have $\psi(c,j)=-\psi(j,c)$. Thus the game is symmetric, its value is zero, and it has an optimal mixed strategy $\cP$ that is used by both players. Optimality of $\cP$ means that $\E_{j\sim\cP}\psi(c,j)\le0$ for every candidate $c$.

For a voter $v$ and a candidate $c$, define the lower and upper ranks of $c$ with respect to $\cP$ as
\[
\rho^-_v(c)=\Pr_{j\sim\cP}[c\succ_v j],\qquad \rho^+_v(c)=\Pr_{j\sim\cP}[c\succeq_v j]=\rho_v^-(c)+\cP(c).
\]
For every candidate $c$,
\begin{align*}
0&\ge\E_{j\sim\cP}\psi(c,j) &&\text{(optimality of $\cP$)}\\
&=\E_{j\sim\cP}\,\mu\big(\{v:c\succ_v j\}\big)+\tfrac12\cP(c)-\tfrac12-\gamma\phi(c)+\gamma\,\E_{\cP}\phi\\
&=\E_v\,\rho^-_v(c)+\tfrac12\cP(c)-\tfrac12-\gamma\phi(c)+\gamma\,\E_{\cP}\phi\\
&\ge\E_v\,\rho^-_v(c)-\tfrac12-\gamma\phi(c)+\gamma\,\E_{\cP}\phi.
\end{align*}
Rearranging, we obtain for every candidate $c$
\begin{equation}\label{eq:eqgame}
\E_v\,\rho^-_v(c)-\gamma\phi(c)\le\frac12-\gamma\,\E_{\cP}\phi.
\end{equation}

Next, fix a voter $v$ and sample $a\sim\cP$. We claim that $\rho^+_v(a)$ stochastically dominates a uniform random variable on $[0,1]$. Indeed, list the candidates from the least preferred to the most preferred by $v$, and assign to each candidate $j$ a subinterval of $[0,1]$ of length $\cP(j)$ in this order. The interval of $j$ is $[\rho^-_v(j),\rho^+_v(j)]$. If we sample a uniform number $w\in[0,1]$ and let $a$ be the candidate whose interval contains $w$, then $a$ is distributed according to $\cP$ and $\rho^+_v(a)\ge w$. Consequently, if $a,b\sim\cP$ are independent and obtained in this way from independent uniform numbers $w_1,w_2\in[0,1]$, then
\begin{align*}
\E_{a,b}\max\{\rho^+_v(a),\rho^+_v(b)\}&\ge\E\max\{w_1,w_2\} &&\text{(since $\rho^+_v(a)\ge w_1$ and $\rho^+_v(b)\ge w_2$)}\\
&=\tfrac23 &&\text{(maximum of two independent uniforms)}.
\end{align*}
Averaging over the voters, and using $\E_{a,b}\bar\phi=\E_{\cP}\phi$, we obtain
\begin{align*}
\E_{a,b}\Big[\E_v\max\{\rho^+_v(a),\rho^+_v(b)\}-\gamma\bar\phi\Big]&=\E_v\,\E_{a,b}\max\{\rho^+_v(a),\rho^+_v(b)\}-\gamma\,\E_{\cP}\phi\\
&\ge\tfrac23-\gamma\,\E_{\cP}\phi.
\end{align*}
Hence we can fix a pair $a,b$ such that
\begin{equation}\label{eq:upperrank}
\E_v\max\{\rho^+_v(a),\rho^+_v(b)\}-\gamma\bar\phi\ge\frac23-\gamma\,\E_{\cP}\phi.
\end{equation}

We now prove the two claims for this pair. Let $c^*$ be a candidate of minimum penalty. Then
\begin{align*}
-\gamma\phi(c^*)&\le\E_v\,\rho^-_v(c^*)-\gamma\phi(c^*)\\
&\le\tfrac12-\gamma\,\E_{\cP}\phi &&\text{(by \eqref{eq:eqgame})},
\end{align*}
that is, $\gamma\,\E_{\cP}\phi\le\gamma\phi(c^*)+\frac12$. Therefore,
\begin{align*}
\gamma\bar\phi&\le\E_v\max\{\rho^+_v(a),\rho^+_v(b)\}-\tfrac23+\gamma\,\E_{\cP}\phi &&\text{(by \eqref{eq:upperrank})}\\
&\le\tfrac13+\gamma\,\E_{\cP}\phi\\
&\le\tfrac13+\gamma\phi(c^*)+\tfrac12.
\end{align*}
Dividing by $\gamma$ gives \eqref{eq:paircost}.

For \eqref{eq:pairdom}, fix a candidate $c$. If $v$ prefers $c$ to both $a$ and $b$, then every candidate that $v$ ranks weakly below $a$ (or below $b$) is ranked strictly below $c$, and hence $\rho^-_v(c)\ge\max\{\rho^+_v(a),\rho^+_v(b)\}$. For all other voters, $\rho^-_v(c)\ge0\ge\max\{\rho^+_v(a),\rho^+_v(b)\}-1$. Let $Q$ be the set of voters that prefer $c$ to both $a$ and $b$, so that $\mu(Q)=q(c)$, and write $\rho^*=\E_v\max\{\rho^+_v(a),\rho^+_v(b)\}$. Then
\begin{align*}
\E_v\,\rho^-_v(c)&\ge\E_v\Big[\max\{\rho^+_v(a),\rho^+_v(b)\}-\mathbf 1_{v\notin Q}\Big] &&\text{(the two pointwise bounds above)}\\
&=\rho^*-\big(1-q(c)\big).
\end{align*}
Therefore,
\begin{align*}
\rho^*-\big(1-q(c)\big)-\gamma\phi(c)&\le\E_v\,\rho^-_v(c)-\gamma\phi(c)\\
&\le\tfrac12-\gamma\,\E_{\cP}\phi &&\text{(by \eqref{eq:eqgame})}\\
&\le \rho^*-\gamma\bar\phi-\tfrac16 &&\text{(by \eqref{eq:upperrank})}.
\end{align*}
Subtracting $\rho^*$ from both sides and rearranging gives \eqref{eq:pairdom}.
\end{proof}

\subsubsection*{Logarithmic Penalties and Inverse-Deficiency Weights}

We apply \cref{lem:pair} with a penalty that is logarithmic in a deficiency-like quantity. Throughout the paper, a \emph{deficiency profile} is a function $\delta:C\to\R_{\ge0}$ such that $\delta(c)\ge\sigma(c)$ for every candidate $c$. (The simplest example is $\delta=\sigma$; in \cref{sec:descent} we use a stronger notion.) Given a deficiency profile $\delta$ and a scale $\xi\in(0,1)$ such that $\min_c\delta(c)\le\xi$, define
\begin{equation}\label{eq:logpenalty}
\lambda=\log(1/\xi),\qquad \hdelta(c)=\delta(c)+\xi,\qquad \phi(c)=\frac{\log\big(\hdelta(c)/\xi\big)}{\lambda},\qquad \gamma=12.
\end{equation}
Notice that $\phi\ge0$, that $\min_c\phi(c)\le\log2/\lambda$, and that $\phi$ is increasing in $\delta$. Indeed, $\hdelta(c)\ge\xi$ gives $\phi\ge0$, and a candidate $c$ with $\delta(c)\le\xi$ has $\hdelta(c)\le2\xi$, and hence $\phi(c)\le\log2/\lambda$. The main property of this penalty is that multiplying $\hdelta$ by a factor of $\eta$ increases the penalty by only $\log \eta/\lambda$: if $\hdelta(c)\le \eta\,\hdelta(a)$, then
\begin{equation}\label{eq:logrepair}
\phi(c)-\phi(a)\le\frac{\log \eta}{\lambda}.
\end{equation}
Indeed,
\begin{align*}
\phi(c)-\phi(a)&=\frac{\log\big(\hdelta(c)/\xi\big)-\log\big(\hdelta(a)/\xi\big)}{\lambda}\\
&=\frac{\log\big(\hdelta(c)/\hdelta(a)\big)}{\lambda}\\
&\le\frac{\log \eta}{\lambda}.
\end{align*}

\begin{definition}[Penalized pair]\label{def:penpair}
Let $\delta$ be a deficiency profile and let $\xi\in(0,1)$ satisfy $\min_c\delta(c)\le\xi$. The \emph{penalized pair} for $(\delta,\xi)$ is the pair $a,b$ given by \cref{lem:pair} with the penalty and the weight $\gamma$ of \eqref{eq:logpenalty}, together with the lottery that elects $a$ and $b$ with probabilities
\begin{equation}\label{eq:inverse}
p_a=\frac{\hdelta(b)}{\hdelta(a)+\hdelta(b)},\qquad p_b=\frac{\hdelta(a)}{\hdelta(a)+\hdelta(b)}.
\end{equation}
(If $a=b$, the lottery simply elects $a$.)
\end{definition}

The weights in \eqref{eq:inverse} are inversely proportional to the deficiencies, so that $p_a\hdelta(a)=p_b\hdelta(b)$. This prevents the candidate with the smaller deficiency from being analyzed at the scale of the larger one.

\begin{lemma}\label{lem:pairsmall}
If $\lambda\ge100$, then the penalized pair satisfies $\hdelta(a),\hdelta(b)\le\xi^{3/4}$.
\end{lemma}
\begin{proof}
We have
\begin{align*}
\phi(a)&\le\phi(a)+\phi(b)\\
&=2\bar\phi\\
&\le2\min_c\phi(c)+\frac{5}{3\gamma} &&\text{(by \eqref{eq:paircost})}\\
&\le\frac{2\log2}{\lambda}+\frac{5}{36} &&\text{(since $\min_c\phi(c)\le\log2/\lambda$ and $\gamma=12$)}\\
&<\frac14 &&\text{(since $\lambda\ge100$)}.
\end{align*}
By the definition of $\phi$, this means $\log(\hdelta(a)/\xi)<\lambda/4=\log(\xi^{-1/4})$, that is, $\hdelta(a)<\xi\cdot\xi^{-1/4}=\xi^{3/4}$. The same holds for $b$.
\end{proof}

The next lemma shows that the inverse-deficiency weights force both candidates of a penalized pair to be individually near-tight in a bad metric.

\begin{lemma}[Individual near-tightness]\label{lem:individual}
Let $a,b$ be a penalized pair for $(\delta,\xi)$ with $\hdelta(a),\hdelta(b)\le\xi^{3/4}\le1/5$, and let $0<\eps\le\xi$. Suppose that some normalized consistent metric satisfies $p_a\SC(a)+p_b\SC(b)\ge3-\eps$. Then for both $i\in\{a,b\}$, in this metric,
\begin{align}
&\SC(i)\ge3-7\hdelta(i),\qquad 2-7\hdelta(i)\le x_i\le5,\label{eq:individual}\\
&\int_0^\infty\pos{g_i(\ell)}\,d\ell\le12\hdelta(i),\qquad \E_v(2y_v-z_v)\le12\hdelta(i).\label{eq:individualslack}
\end{align}
\end{lemma}
\begin{proof}
If $a=b$, then $\SC(a)\ge3-\eps\ge3-\hdelta(a)$. Otherwise, notice that $\sigma(i)\le\delta(i)\le\hdelta(i)\le1/5$ for both $i$. Then
\begin{align*}
\SC(b)&\le3+\frac{4\sigma(b)}{1-\sigma(b)} &&\text{(\cref{lem:cost})}\\
&\le3+5\sigma(b) &&\text{(since $\sigma(b)\le1/5$)}\\
&\le3+5\delta(b).
\end{align*}
Therefore,
\begin{align*}
p_a\SC(a)&\ge3-\eps-p_b\SC(b) &&\text{(assumption on the metric)}\\
&\ge3-\eps-p_b\big(3+5\delta(b)\big) &&\text{(bound on $\SC(b)$ above)}\\
&=3p_a-\eps-5p_b\,\delta(b).
\end{align*}
Dividing by $p_a>0$,
\begin{align*}
3-\SC(a)&\le\frac{\eps}{p_a}+5\,\frac{p_b}{p_a}\,\delta(b)\\
&=\eps\Big(1+\frac{\hdelta(a)}{\hdelta(b)}\Big)+5\,\frac{\hdelta(a)}{\hdelta(b)}\,\delta(b) &&\text{(by \eqref{eq:inverse})}\\
&=\eps+\hdelta(a)\cdot\frac{\eps}{\hdelta(b)}+5\,\hdelta(a)\cdot\frac{\delta(b)}{\hdelta(b)}\\
&\le\hdelta(a)+\hdelta(a)+5\,\hdelta(a) &&\text{(since $\eps\le\xi\le\hdelta(a)$ and $\eps,\delta(b)\le\hdelta(b)$)}\\
&=7\,\hdelta(a).
\end{align*}
The argument for $b$ is symmetric. For both $i\in\{a,b\}$, \cref{lem:cost} also gives $x_i\le4/(1-\sigma(i))\le5$, since $\sigma(i)\le1/5$. Moreover,
\begin{align*}
x_i&\ge\E_v e_v(i) &&\text{(since $e_v(i)\le x_i$ by \cref{lem:profile})}\\
&\ge\SC(i)-1 &&\text{(\cref{lem:profile})}\\
&\ge2-7\hdelta(i) &&\text{(bound on $\SC(i)$ above)}.
\end{align*}
Finally, the second part of \cref{lem:cost} with $\alpha=7\hdelta(i)$ gives
\begin{align*}
\int_0^\infty\pos{g_i(\ell)}\,d\ell&\le7\hdelta(i)+\sigma(i)\,x_i &&\text{(\cref{lem:cost})}\\
&\le7\hdelta(i)+5\,\sigma(i) &&\text{(since $x_i\le5$)}\\
&\le12\,\hdelta(i),
\end{align*}
and the same bound holds for $\E_v(2y_v-z_v)$.
\end{proof}

%% file: src/blocks.tex
\subsection{Extracting a Fuzzy Block}\label{sec:blocks}

In this subsection, we show how to extract a fuzzy block (\cref{def:block}) from a bad metric.

The following lemma extracts a block from a metric in which the top choices of the voters are almost as far from the optimum as possible. It chooses the three thresholds at widely separated scales, inside a range of distances that contains few top choices, and with an inner threshold at which given slack functions are small.

\begin{lemma}[Threshold extraction]\label{lem:threshold}
Let $0<\tau\le10^{-10}$, let $0<\xi\le\exp(-\tau^{-16})$, and let $\lambda=\log(1/\xi)$. Fix a normalized consistent metric with $\E_v(2y_v-z_v)\le100\,\xi^{3/4}$, and let $f_1,f_2$ be nonnegative integrable functions on $[0,\infty)$ with $\int_0^\infty f_i(\ell)\,d\ell\le\beta_i$ for some $\beta_i\le100\,\xi^{3/4}$. Then there are numbers $\zeta\in[e^{-\sqrt\lambda},1)$ and $\ell_0<\ell_1<\ell_2$ such that $(I_{\ell_0},I_{\ell_1},I_{\ell_2})$ is a $\tau$-block, and
\[
\ell_0\in[\zeta,2\zeta],\qquad \ell_2\le\tau^2,\qquad f_i(\ell_0)\le\frac{4\beta_i}{\zeta}\le\tau^2\ (i=1,2),\qquad \frac{12\log(52/\zeta)}{\lambda}\le\tau^2.
\]
\end{lemma}
\begin{proof}
Notice that $\lambda\ge\tau^{-16}\ge10^{160}$. Divide the interval $[e^{-\sqrt\lambda},e^{-\lambda^{1/4}}]$ into $h=\lfloor\lambda^{1/8}\rfloor$ intervals whose endpoints form a geometric progression. Since the voters have total mass one, one of these intervals, say $(\zeta,\zeta']$, satisfies $\mu(\{v:\zeta<z_v\le \zeta'\})\le1/h$. Since $\lambda^{1/8}\ge\tau^{-2}\ge1$,
\begin{align*}
\frac1h&\le\frac{2}{\lambda^{1/8}}\\
&\le2\tau^2.
\end{align*}
Moreover,
\begin{align*}
\log(\zeta'/\zeta)&=\frac{\sqrt\lambda-\lambda^{1/4}}{h}\\
&\ge\frac{\sqrt\lambda-\lambda^{1/4}}{\lambda^{1/8}}\\
&=\lambda^{3/8}-\lambda^{1/8}\\
&\ge\frac{\lambda^{3/8}}{2},
\end{align*}
and therefore $\sqrt{\zeta'/\zeta}=\exp\big(\tfrac12\log(\zeta'/\zeta)\big)\ge\exp(\lambda^{3/8}/4)>8$.

By Markov's inequality, the set of $\ell\in[\zeta,2\zeta]$ with $f_i(\ell)>4\beta_i/\zeta$ has length at most $\zeta/4$, for each $i$ (if $\beta_i=0$ then $f_i=0$ almost everywhere). Indeed, its length is at most $\int_0^\infty f_i(\ell)\,d\ell\big/(4\beta_i/\zeta)\le\beta_i/(4\beta_i/\zeta)=\zeta/4$. Hence we can choose $\ell_0\in[\zeta,2\zeta]$ with $f_i(\ell_0)\le4\beta_i/\zeta$ for both $i$. Set $\ell_1=\sqrt{\zeta\zeta'}$ and $\ell_2=\zeta'$. Then
\begin{align*}
\ell_1&=\sqrt{\zeta'/\zeta}\cdot \zeta\\
&>8\zeta &&\text{(since $\sqrt{\zeta'/\zeta}>8$)}\\
&\ge4\ell_0,
\end{align*}
and $\ell_2/\ell_1=\sqrt{\zeta'/\zeta}>8$. Let $I_0=I_{\ell_0}$, $I_1=I_{\ell_1}$, $I_2=I_{\ell_2}$, and $G=G_{I_0}$.

Since $\ell_0\ge \zeta$ and $\ell_2=\zeta'$, every voter with top choice in $I_2\setminus I_0$ satisfies $\zeta<z_v\le \zeta'$. Therefore $\plu(I_2\setminus I_0)\le2\tau^2$.

Now consider a voter $v\in G$, so that $z_v\le\ell_0$. Since $2y_v-z_v\ge0$ for every voter (\cref{lem:profile}),
\begin{align*}
\mu\big(\{v:2y_v-z_v>\ell_0\}\big)&\le\frac{\E_v(2y_v-z_v)}{\ell_0} &&\text{(Markov's inequality)}\\
&\le\frac{100\,\xi^{3/4}}{\ell_0} &&\text{(assumption of the lemma)}\\
&\le\frac{100\,\xi^{3/4}}{\zeta}.
\end{align*}
Consider a voter $v\in G$ with $2y_v-z_v\le\ell_0$. Then $2y_v\le\ell_0+z_v\le2\ell_0$, that is, $y_v\le\ell_0$. For every candidate $c\in I_0$ and every candidate $j\notin I_1$,
\begin{align*}
d(v,c)&\le d(v,o)+d(o,c)\\
&=y_v+x_c\\
&\le2\ell_0 &&\text{(since $y_v\le\ell_0$ and $c\in I_{\ell_0}$)},\\
d(v,j)&\ge d(o,j)-d(o,v)\\
&=x_j-y_v\\
&>\ell_1-\ell_0 &&\text{(since $j\notin I_{\ell_1}$ and $y_v\le\ell_0$)}\\
&\ge3\ell_0 &&\text{(since $\ell_1\ge4\ell_0$)}.
\end{align*}
Hence $d(v,c)<d(v,j)$, and consistency gives $c\succ_v j$. Thus $v$ ranks $I_0$ above $C\setminus I_1$. Similarly, for every $c\in I_1$ and every $j\notin I_2$,
\begin{align*}
d(v,c)&\le y_v+x_c\\
&\le\ell_0+\ell_1 &&\text{(since $y_v\le\ell_0$ and $c\in I_{\ell_1}$)},\\
d(v,j)&\ge x_j-y_v\\
&>\ell_2-\ell_0 &&\text{(since $j\notin I_{\ell_2}$ and $y_v\le\ell_0$)}\\
&>8\ell_1-\ell_0 &&\text{(since $\ell_2>8\ell_1$)}\\
&>\ell_1+\ell_0,
\end{align*}
so $v$ ranks $I_1$ above $C\setminus I_2$.

It remains to verify the numerical bounds. We have $\zeta\ge e^{-\sqrt\lambda}$ and $\zeta<\zeta'\le e^{-\lambda^{1/4}}<1$. Recall that $\xi^{3/4}=e^{-3\lambda/4}$ and $\lambda\ge\tau^{-16}\ge10^{160}$. The exceptional mass satisfies
\begin{align*}
\frac{100\,\xi^{3/4}}{\zeta}&\le100\,e^{-3\lambda/4+\sqrt\lambda} &&\text{(since $\zeta\ge e^{-\sqrt\lambda}$)}\\
&\le100\,e^{-\lambda/2}\\
&\le\tau^2.
\end{align*}
Similarly,
\begin{align*}
\frac{4\beta_i}{\zeta}&\le\frac{400\,\xi^{3/4}}{\zeta}\\
&\le400\,e^{-\lambda/2}\\
&\le\tau^2.
\end{align*}
Also,
\begin{align*}
\ell_2&=\zeta'\le e^{-\lambda^{1/4}}\\
&\le e^{-\tau^{-4}}\\
&\le\tau^2.
\end{align*}
Finally,
\begin{align*}
\frac{12\log(52/\zeta)}{\lambda}&\le\frac{12\big(\log52+\sqrt\lambda\big)}{\lambda} &&\text{(since $\zeta\ge e^{-\sqrt\lambda}$)}\\
&\le\frac{24}{\sqrt\lambda}\\
&\le24\,\tau^{8}\\
&\le\tau^2.
\end{align*}
\end{proof}

We now apply the threshold extraction to a bad metric of a penalized pair. The resulting lemma directly gives \cref{lem:pairtoblock}.

\begin{lemma}[Blocks from a penalized pair]\label{lem:pairblock}
Let $0<\tau\le10^{-10}$, let $\delta$ be a deficiency profile, and let $\xi$ and $\eps$ satisfy $0<\eps\le\xi\le\exp(-\tau^{-16})$ and $\min_c\delta(c)\le\xi$. Let $a,b$ be the penalized pair for $(\delta,\xi)$, and suppose that some normalized consistent metric satisfies $p_a\SC(a)+p_b\SC(b)\ge3-\eps$. Then there exist a $\tau$-block $(J_0,J_1,J_2)$ and a number $\zeta\in(0,1)$ with $12\log(52/\zeta)/\lambda\le\tau^2$ such that:
\begin{enumerate}
\item $a,b\notin J_2$;
\item for both $i\in\{a,b\}$, we have $-\delta(i)\le r_{J_0}-t_{J_0}(i)\le 48\,\hdelta(i)/\zeta\le\tau^2$, and $\delta(i)\le\hdelta(i)\le\tau^2$;
\item the mass of the block satisfies $s_{J_0}\le3/4$;
\item if $a'\in\{a,b\}$ has the smaller deficiency $\delta(a')=\min\{\delta(a),\delta(b)\}$, and some candidate $c\in J_0$ satisfies $\hdelta(c)\le50\,\hdelta(a')/\zeta$, then $s_{J_0}\ge1/20$.
\end{enumerate}
\end{lemma}
\begin{proof}
Since $\lambda\ge\tau^{-16}\ge100$, \cref{lem:pairsmall} gives $\hdelta(a),\hdelta(b)\le\xi^{3/4}\le1/5$, and \cref{lem:individual} applies. In particular, $\E_v(2y_v-z_v)\le12\,\xi^{3/4}$. Apply \cref{lem:threshold} with $f_1=\pos{g_a}$, $f_2=\pos{g_b}$, and $\beta_i=12\hdelta(i)\le12\,\xi^{3/4}$, and let $J_k=I_{\ell_k}$ for $k=0,1,2$.

For (i), let $i\in\{a,b\}$. Then
\begin{align*}
x_i&\ge2-7\hdelta(i) &&\text{(\cref{lem:individual})}\\
&>\tau^2 &&\text{(since $\hdelta(i)\le1/5$)}\\
&\ge\ell_2 &&\text{(\cref{lem:threshold})},
\end{align*}
so $i\notin I_{\ell_2}=J_2$. For (ii), notice that $g_i(\ell_0)=r_{J_0}-t_{J_0}(i)$. For the upper bound,
\begin{align*}
r_{J_0}-t_{J_0}(i)&\le\pos{g_i(\ell_0)}\\
&=f_i(\ell_0)\le\frac{4\beta_i}{\zeta} &&\text{(\cref{lem:threshold})}\\
&=\frac{48\,\hdelta(i)}{\zeta}\\
&\le\tau^2 &&\text{(\cref{lem:threshold})}.
\end{align*}
The lower bound follows from \cref{obs:cut}, which gives $g_i(\ell_0)\ge-\sigma(i)\ge-\delta(i)$. Also $\delta(i)\le\hdelta(i)\le\xi^{3/4}\le\tau^2$.

For (iii), consider $i=a$. Every voter $v\notin T_{J_0}(a)$ ranks some candidate $j\in J_0$ weakly below $a$, so $e_v(a)\le x_j\le\ell_0$; every other voter has $e_v(a)\le x_a\le5$. Hence
\begin{align*}
2-7\hdelta(a)&\le\SC(a)-1 &&\text{(\cref{lem:individual})}\\
&\le\E_ve_v(a) &&\text{(\cref{lem:profile})}\\
&\le\ell_0\big(1-t_{J_0}(a)\big)+5\,t_{J_0}(a) &&\text{(the two bounds on $e_v(a)$ above)}\\
&\le\ell_0+5\,t_{J_0}(a).
\end{align*}
Therefore,
\begin{align*}
t_{J_0}(a)&\ge\frac{2-7\hdelta(a)-\ell_0}{5}\\
&\ge\frac{2-8\tau^2}{5} &&\text{(since $\hdelta(a)\le\tau^2$ by (ii), and $\ell_0\le\ell_2\le\tau^2$)}\\
&\ge0.39,
\end{align*}
and
\begin{align*}
s_{J_0}&=1-r_{J_0}\\
&\le1-t_{J_0}(a)+\delta(a) &&\text{(lower bound in (ii))}\\
&\le1-0.39+\tau^2 &&\text{(by (ii))}\\
&\le3/4.
\end{align*}

For (iv), the penalty is increasing in $\delta$, so $\phi(a')=\min\{\phi(a),\phi(b)\}\le\bar\phi$. Since $\hdelta(c)\le(50/\zeta)\,\hdelta(a')$,
\begin{align*}
\gamma\big(\phi(c)-\bar\phi\big)&\le\gamma\big(\phi(c)-\phi(a')\big)\\
&\le\frac{12\log(50/\zeta)}{\lambda} &&\text{(by \eqref{eq:logrepair} with $\eta=50/\zeta$, and $\gamma=12$)}\\
&\le\frac{12\log(52/\zeta)}{\lambda}\\
&\le\tau^2 &&\text{(choice of $\zeta$)}.
\end{align*}
Since $c\in J_0$, every voter of $T_{J_0}(a)\cap T_{J_0}(b)$ prefers $c$ to both $a$ and $b$. By (ii), $t_{J_0}(i)\ge r_{J_0}-\tau^2$ for both $i$. Hence
\begin{align*}
q(c)&\ge\mu\big(T_{J_0}(a)\cap T_{J_0}(b)\big) &&\text{(these voters prefer $c$ to $a$ and $b$)}\\
&\ge t_{J_0}(a)+t_{J_0}(b)-1\\
&\ge2\big(r_{J_0}-\tau^2\big)-1 &&\text{(by (ii))}\\
&=1-2s_{J_0}-2\tau^2.
\end{align*}
Plugging these bounds into \eqref{eq:pairdom},
\begin{align*}
\tfrac16&\le1-q(c)+\gamma\big(\phi(c)-\bar\phi\big) &&\text{(by \eqref{eq:pairdom})}\\
&\le1-\big(1-2s_{J_0}-2\tau^2\big)+\tau^2\\
&=2s_{J_0}+3\tau^2.
\end{align*}
Hence $s_{J_0}\ge1/12-3\tau^2/2\ge1/20$.
\end{proof}

\subsection{Proof of \Cref{lem:pairtoblock}}\label{sec:p2bproof}

\gdef\ptbnote{}\pairtoblocklemma*
\begin{proof}
A function $\delta$ with $\delta\ge\sigma$ is a deficiency profile. Let $a,b$ be the penalized pair for $(\delta,\xi)$ (\cref{def:penpair}), and let $p=p_a$. Suppose that the lottery that elects $a$ with probability $p$ and $b$ with probability $1-p$ has distortion larger than $3-\eps$. Then it has a bad metric (\cref{sec:profiles}), and \cref{lem:pairblock} gives a $\tau$-block $(J_0,J_1,J_2)$ and a number $\zeta\in(0,1)$. Since $\hdelta=\delta+\xi$, properties (i)--(iv) of \cref{lem:pairblock} are exactly properties (i)--(iv) of \cref{lem:pairtoblock}.
\end{proof}

%% file: src/coupled.tex
\section{Proof of the Protected Matching Lemma}\label{sec:coupled}

In this section, we prove \cref{lem:protected}. The proof uses a joint selection procedure whose two output candidates share a common distortion certificate. From a block and a bad metric for the output of the procedure, we construct a candidate with a coupling that has tiny deficiency and keeps a constant amount of mass inside the block.

\subsection{The Coupled Veto Process}

Let $A$ and $B=C\setminus A$ be a partition of the candidates with $s=\plu(A)\in(0,1)$. We run two veto processes simultaneously, one on $A$ and one on $B$. Every candidate starts with a score equal to its plurality share, so the total initial score is $s$ in $A$ and $1-s$ in $B$. At time $t\in[0,1)$, let $a_v(t)$ and $b_v(t)$ be the least preferred candidates of voter $v$ among the candidates with positive score in $A$ and in $B$, respectively. Let $\eta_v(t)=1$ if $b_v(t)\succ_v a_v(t)$ and $\eta_v(t)=-1$ otherwise, and let $\bar\eta(t)=\E_v\eta_v(t)$. Set
\begin{equation}\label{eq:couple}
\theta=\tfrac14\min\{s,1-s\},\qquad \omega_v(t)=s+\theta\big(\eta_v(t)-\bar\eta(t)\big).
\end{equation}
Voter $v$ decreases the score of $a_v(t)$ at rate $\omega_v(t)/n$ and the score of $b_v(t)$ at rate $(1-\omega_v(t))/n$. In words, every voter vetoes slightly more on the side where its worst remaining candidate is worse.

The total rate in $A$ is exactly 
\begin{align*}
\E_v\,\omega_v(t)&=s+\theta\big(\E_v\eta_v(t)-\bar\eta(t)\big)\\
&=s,
\end{align*}and the total rate in $B$ is exactly $1-s$. Therefore both processes run until time one, when all scores are exhausted. Between two consecutive times at which some score is exhausted, the candidates $a_v(t)$ and $b_v(t)$, and hence all rates, are fixed, so all scores decrease linearly. Therefore the process is well-defined. Every voter spends a constant fraction of its rate on each side: since $|\eta_v(t)-\bar\eta(t)|\le2$ and $2\theta\le\frac12\min\{s,1-s\}$,
\begin{equation}\label{eq:ratebounds}
\omega_v(t)\ge s-2\theta\ge s/2\qquad\text{and}\qquad 1-\omega_v(t)\ge1-s-2\theta\ge(1-s)/2.
\end{equation}
Let $a\in A$ and $b\in B$ be candidates whose scores are exhausted last in their processes, and let
\[
p=\int_0^1\big(s-\theta\bar\eta(t)\big)\,dt\in[s-\theta,s+\theta],
\]
where the inclusion holds since $|\bar\eta(t)|\le1$.
The output of the \emph{coupled veto} procedure is the two-candidate lottery that elects $a$ with probability $p$ and $b$ with probability $1-p$.

For a set $J\subseteq C$, define the \emph{mixed cut value} and the \emph{disagreement} of this lottery as
\[
\tbar_J=p\,t_J(a)+(1-p)\,t_J(b),\qquad \nu_J=\mu\big(T_J(a)\,\triangle\,T_J(b)\big).
\]
Moreover, let $\chi_J$ be the total veto mass that the voters outside $T_J(a)\cup T_J(b)$ direct at candidates outside $J$ (over both processes and the whole time interval).

\begin{lemma}[Coupled veto certificate]\label{lem:coupled}
For every set $J\subseteq C$,
\[
\tbar_J+\theta\,\nu_J+\chi_J\le r_J.
\]
Consequently, the output of the coupled veto procedure has distortion at most $3$.
\end{lemma}
\begin{proof}
The candidates outside $J$ have total initial score $r_J$, and all of it is vetoed by time one. We lower bound the veto mass that every voter directs outside $J$. Since $a$ and $b$ have positive score at every time $t<1$, we have $a\succeq_v a_v(t)$ and $b\succeq_v b_v(t)$ for every voter $v$.

Consider a voter $v\in T_J(a)\cap T_J(b)$. It ranks all of $J$ above $a$ and above $b$, so neither $a_v(t)$ nor $b_v(t)$ lies in $J$, and its whole veto mass of $1/n$ goes outside $J$.

Next, consider a voter $v\in T_J(a)\setminus T_J(b)$. Again $a_v(t)\notin J$. If $b_v(t)\in J$, then $b_v(t)\succ_v a\succeq_v a_v(t)$, so $\eta_v(t)=1$ and the rate of $v$ outside $J$ is $\omega_v(t)/n=(s+\theta-\theta\bar\eta(t))/n$. If $b_v(t)\notin J$, the rate of $v$ outside $J$ is $1/n$, which is larger. Integrating over time, $v$ directs outside $J$ a mass of at least
\begin{align*}
\frac1n\int_0^1\big(s+\theta-\theta\bar\eta(t)\big)\,dt&=\frac1n\Big(\theta+\int_0^1\big(s-\theta\bar\eta(t)\big)\,dt\Big)\\
&=\frac{p+\theta}{n}.
\end{align*} Symmetrically, a voter $v\in T_J(b)\setminus T_J(a)$ directs a mass of at least $(1-p+\theta)/n$ outside $J$, since in the critical case $\eta_v(t)=-1$ and its rate outside $J$ is $(1-\omega_v(t))/n=(1-s+\theta+\theta\bar\eta(t))/n$.

Finally, by definition, the voters outside $T_J(a)\cup T_J(b)$ direct a mass of $\chi_J$ outside $J$. Write $T^a=T_J(a)$ and $T^b=T_J(b)$. Summing up,
\begin{align*}
r_J&\ge\mu(T^a\cap T^b)+(p+\theta)\,\mu(T^a\setminus T^b)+(1-p+\theta)\,\mu(T^b\setminus T^a)+\chi_J\\
&=\Big(p\,\mu(T^a)+(1-p)\,\mu(T^b)\Big)+\theta\Big(\mu(T^a\setminus T^b)+\mu(T^b\setminus T^a)\Big)+\chi_J\\
&=\tbar_J+\theta\,\nu_J+\chi_J.
\end{align*}

For the distortion bound, fix a normalized consistent metric and apply the inequality to $J=I_\ell$ for every $\ell\ge0$. Then
\begin{align*}
p\,\SC(a)+(1-p)\,\SC(b)&\le1+p\,\E_ve_v(a)+(1-p)\,\E_ve_v(b) &&\text{(\cref{lem:profile})}\\
&=1+\int_0^\infty\tbar_{I_\ell}\,d\ell &&\text{(\cref{lem:profile})}\\
&\le1+\int_0^\infty r_{I_\ell}\,d\ell &&\text{(\cref{lem:coupled})}\\
&=1+\E_v z_v &&\text{(\cref{lem:profile})}\\
&\le3 &&\text{(\cref{lem:profile})}.\qedhere
\end{align*}
\end{proof}

\subsection{Protected Matchings}

We now restate and prove \cref{lem:protected}.

\gdef\pmnote{}\protectedlemma*
\begin{proof}
Suppose that every two-candidate lottery has distortion larger than $3-\eps$; we show that the second alternative holds. Run the coupled veto procedure with $A=I_0$ and $B=C\setminus I_0$. Its output is a two-candidate lottery, so it has a bad metric (\cref{sec:profiles}); fix this metric. Notice that $\theta=\frac14\min\{s,1-s\}\ge\frac14\cdot\frac1{20}=\frac1{80}$, since $s\in[1/20,3/4]$. Then
\begin{align*}
2-\eps&\le p\big(\SC(a)-1\big)+(1-p)\big(\SC(b)-1\big) &&\text{(the metric is bad)}\\
&\le p\,\E_ve_v(a)+(1-p)\,\E_ve_v(b) &&\text{(\cref{lem:profile})}\\
&=\int_0^\infty\tbar_{I_\ell}\,d\ell &&\text{(\cref{lem:profile})}\\
&\le\int_0^\infty\big(r_{I_\ell}-\theta\,\nu_{I_\ell}\big)\,d\ell &&\text{(\cref{lem:coupled}, with $\chi_{I_\ell}\ge0$)}\\
&=\E_vz_v-\theta\int_0^\infty\nu_{I_\ell}\,d\ell &&\text{(\cref{lem:profile})}.
\end{align*}
We draw three consequences. First,
\begin{align*}
\E_v(2y_v-z_v)&=2-\E_vz_v\\
&\le2-(2-\eps) &&\text{(chain above, and $\theta\int_0^\infty\nu_{I_\ell}\,d\ell\ge0$)}\\
&=\eps.
\end{align*}
Second,
\begin{align*}
\int_0^\infty\big(r_{I_\ell}-\tbar_{I_\ell}\big)\,d\ell&=\E_vz_v-\int_0^\infty\tbar_{I_\ell}\,d\ell &&\text{(\cref{lem:profile})}\\
&\le2-(2-\eps) &&\text{(\cref{lem:profile} and the chain above)}\\
&=\eps.
\end{align*}
Third,
\begin{align*}
\int_0^\infty\nu_{I_\ell}\,d\ell&\le\frac{\E_vz_v-(2-\eps)}{\theta}\\
&\le\frac{\eps}{\theta} &&\text{(\cref{lem:profile})}\\
&\le80\,\eps &&\text{(since $\theta\ge1/80$)}.
\end{align*}
The function $r_{I_\ell}-\tbar_{I_\ell}$ is nonnegative by \cref{lem:coupled}. Moreover,
\begin{align*}
\big|\E_ve_v(a)-\E_ve_v(b)\big|&\le\E_v\big|e_v(a)-e_v(b)\big| &&\text{(Jensen's inequality)}\\
&=\int_0^\infty\nu_{I_\ell}\,d\ell &&\text{(\cref{lem:profile})}\\
&\le80\,\eps &&\text{(third consequence)}.
\end{align*}
Hence
\begin{align*}
\E_ve_v(a)&\ge p\,\E_ve_v(a)+(1-p)\,\E_ve_v(b)-(1-p)\big|\E_ve_v(a)-\E_ve_v(b)\big|\\
&\ge(2-\eps)-80\,\eps\\
&=2-81\,\eps,
\end{align*}
and similarly $\E_ve_v(b)\ge2-81\,\eps$.

Apply \cref{lem:threshold} with $\xi=\eps$, $f_1(\ell)=r_{I_\ell}-\tbar_{I_\ell}$, $f_2(\ell)=\nu_{I_\ell}$, and $\beta_1=\beta_2=80\eps$. We obtain a $\tau$-block $(J_0,J_1,J_2)$ with
\begin{equation}\label{eq:protbad}
0\le r_{J_0}-\tbar_{J_0}\le\tau^2\qquad\text{and}\qquad\nu_{J_0}\le\tau^2.
\end{equation}
Moreover, 
\begin{align*}
x_a&\ge\E_ve_v(a) &&\text{(since $e_v(a)\le x_a$ by \cref{lem:profile})}\\
&\ge2-81\,\eps\\
&>\ell_2 &&\text{(since $\ell_2\le\tau^2$ by \cref{lem:threshold})},
\end{align*}
and similarly for $b$, so $a,b\notin J_2$.

\medskip\noindent\textbf{The new block is small.} We claim that $s_{J_0}<1000\tau^2$. Let $V_0=V\setminus(T_{J_0}(a)\cup T_{J_0}(b))$. Write $T^a=T_{J_0}(a)$ and $T^b=T_{J_0}(b)$. Then
\begin{align*}
\mu(T^a\cup T^b)-\tbar_{J_0}&=\mu(T^a\cup T^b)-p\,\mu(T^a)-(1-p)\,\mu(T^b)\\
&=(1-p)\,\mu(T^a\setminus T^b)+p\,\mu(T^b\setminus T^a)\\
&\le\nu_{J_0}.
\end{align*}
Hence
\begin{align*}
\mu(V_0)&=1-\mu(T^a\cup T^b)\\
&\ge1-\tbar_{J_0}-\nu_{J_0}\\
&\ge1-r_{J_0}-\nu_{J_0} &&\text{(\cref{lem:coupled})}\\
&=s_{J_0}-\nu_{J_0}\\
&\ge s_{J_0}-\tau^2 &&\text{(by \eqref{eq:protbad})}.
\end{align*}
Also, the voters of $V_0$ direct a total veto mass of $\chi_{J_0}$ outside $J_0$, and
\begin{align*}
\chi_{J_0}&\le r_{J_0}-\tbar_{J_0}-\theta\,\nu_{J_0} &&\text{(\cref{lem:coupled})}\\
&\le r_{J_0}-\tbar_{J_0}\\
&\le\tau^2 &&\text{(by \eqref{eq:protbad})}.
\end{align*}
By \eqref{eq:ratebounds}, every voter spends at least a $s/2\ge1/40$ fraction of its veto mass in the $A$-process and at least a $(1-s)/2\ge1/8$ fraction in the $B$-process. The total veto mass received by a candidate equals its plurality share. Therefore
\begin{align*}
\plu(J_0\cap A)&\ge\text{(veto mass from $V_0$ to $J_0\cap A$)} &&\text{(veto mass equals plurality share)}\\
&\ge\text{(veto mass from $V_0$ in the $A$-process)}-\chi_{J_0}\\
&\ge\mu(V_0)/40-\tau^2 &&\text{(by \eqref{eq:ratebounds}, and $\chi_{J_0}\le\tau^2$)}.
\end{align*}
Similarly, $\plu(J_0\cap B)\ge\mu(V_0)/8-\tau^2$.

Suppose for contradiction that $s_{J_0}\ge1000\tau^2$. Then $\mu(V_0)\ge s_{J_0}-\tau^2\ge999\tau^2$, and
\begin{align*}
\plu(J_0\cap A)&\ge\frac{999\,\tau^2}{40}-\tau^2>23\,\tau^2, &
\plu(J_0\cap B)&\ge\frac{999\,\tau^2}{8}-\tau^2>23\,\tau^2.
\end{align*} Since $B=C\setminus I_0$ and $\plu(I_2\setminus I_0)\le2\tau^2$, there is a candidate $j\in J_0\setminus I_2$. Moreover, the set $G_{I_0}\cap G_{J_0}=G_{I_0\cap J_0}$ has mass $\plu(J_0\cap A)>23\tau^2$, so it contains a voter $v$ that is a core voter of both blocks. Since $a\in A=I_0$ and $j\notin I_2\supseteq I_1$, the old block gives $a\succ_v j$. Since $j\in J_0$ and $a\notin J_2\supseteq J_1$, the new block gives $j\succ_v a$. This is a contradiction, which proves the claim.

\medskip\noindent\textbf{Building the protected matching.} We build a labeled coupling using the two veto processes. Whenever a voter $v$ decreases the score of a candidate $j$ in the $A$-process, we add the same amount of mass to edges from $v$ to the voters $u$ with $\tp(u)=j$, in equal proportions, and label these edges by $a$. We do the same in the $B$-process with label $b$. Every voter decreases scores at a total rate of $1/n$ during the time interval $[0,1)$, so it sends a total mass of $1/n$. The score of a candidate $j$ starts at $\plu(j)$, is exhausted by time one, and is distributed equally among the $n\,\plu(j)$ voters $u$ with $\tp(u)=j$, so each of these voters receives a total mass of $1/n$. Hence we obtain a labeled coupling $\pi_a,\pi_b$, and $\pi=\pi_a+\pi_b$ is a coupling. If $v$ sends mass to $u$ along an edge labeled $a$ at time $t$, then $\tp(u)=a_v(t)$ is the least preferred candidate of $v$ among the candidates of $A$ with positive score at time $t$, and $a$ is one of these candidates, since its score remains positive until time one. Hence $a\succeq_v\tp(u)$, that is, the edge is valid for $a$. The same argument applies to the edges labeled $b$. So the invalid mass is zero.

All destinations of the $A$-process lie in $G=G_{I_0}$, and by \eqref{eq:ratebounds}, every voter of $G$ spends at least an $s/2$ fraction of its mass in the $A$-process. Hence
\begin{align*}
\pi(G,G)&\ge\pi_a(G,V) &&\text{(all destinations of $\pi_a$ lie in $G$)}\\
&\ge\mu(G)\cdot\frac s2 &&\text{(by \eqref{eq:ratebounds})}\\
&=\frac{s^2}{2}.
\end{align*}

Next, we compare the cut mass of the labeled coupling with $\tbar_{J_0}$. A voter $v$ sends mass $w_v/n$ along edges labeled $a$ and $(1-w_v)/n$ along edges labeled $b$, where $w_v=\int_0^1\omega_v(t)\,dt$. Its contribution to $\tbar_\pi(J_0)$ is $(w_v\mathbf 1_{v\in T_{J_0}(a)}+(1-w_v)\mathbf 1_{v\in T_{J_0}(b)})/n$, while its contribution to $\tbar_{J_0}$ is $(p\,\mathbf 1_{v\in T_{J_0}(a)}+(1-p)\mathbf 1_{v\in T_{J_0}(b)})/n$. The difference of the two contributions is $(w_v-p)\big(\mathbf 1_{v\in T_{J_0}(a)}-\mathbf 1_{v\in T_{J_0}(b)}\big)/n$. Therefore
\begin{align*}
\big|\tbar_\pi(J_0)-\tbar_{J_0}\big|&\le\frac1n\sum_v|w_v-p|\cdot\big|\mathbf 1_{v\in T_{J_0}(a)}-\mathbf 1_{v\in T_{J_0}(b)}\big|\\
&\le\frac1n\sum_v\big|\mathbf 1_{v\in T_{J_0}(a)}-\mathbf 1_{v\in T_{J_0}(b)}\big|\\
&=\nu_{J_0}.
\end{align*}

Finally, apply \cref{lem:repairlabeled} to the labeled coupling and the set $J_0$. We obtain a candidate $c\in J_0$ and a coupling $\pi'$ with

\begin{align*}
\defi_c(\pi')&\le0+r_{J_0}-\tbar_\pi(J_0) &&\text{(\cref{lem:repairlabeled}, with invalid mass zero)}\\
&\le r_{J_0}-\tbar_{J_0}+\nu_{J_0} &&\text{(since $|\tbar_\pi(J_0)-\tbar_{J_0}|\le\nu_{J_0}$)}\\
&\le2\tau^2 &&\text{(by \eqref{eq:protbad})},
\end{align*}
and
\begin{align*}
\pi'(G,G)&\ge\pi(G,G)-\mu(G_{J_0}\setminus G) &&\text{(\cref{lem:repairlabeled})}\\
&\ge s^2/2-s_{J_0} &&\text{(since $\pi(G,G)\ge s^2/2$ and $\mu(G_{J_0}\setminus G)\le\mu(G_{J_0})=s_{J_0}$)}\\
&\ge s^2/2-1000\tau^2 &&\text{(the new block is small)}\\
&\ge s^2/4 &&\text{(since $s\ge1/20$ and $\tau\le10^{-10}$)}.
\end{align*}
\end{proof}

%% file: src/descent3.tex
\section{Proof of the Initial Block and Descent Lemmas}\label{sec:descent}

In this section, we prove \cref{lem:initial,lem:step}. Besides the definitions of \cref{sec:prelim} and the fuzzy blocks of \cref{def:block}, we use four lemmas from \cref{sec:results}: the weighted veto lemma (\cref{lem:weighted}), the repair lemma (\cref{lem:repair}), the pair-to-block lemma (\cref{lem:pairtoblock}), and the protected matching lemma (\cref{lem:protected}). We first prove \cref{lem:initial}, whose proof is short. We then prove \cref{lem:step}, whose proof is longer: we first fix the setting and prove five auxiliary claims, and then combine them.

Throughout this section, write $\eps=\eps_*$. Notice that $\tau_0<\tau_1<\cdots<\tau_{4000}=10^{-10}$, and hence $\eps\le\exp(-\tau_j^{-16})$ for every $j$. Fix an election.

Each of \cref{lem:initial,lem:step} states that one of two alternatives holds: either some two-candidate lottery has distortion at most $3-\eps$, or a certain block exists. If the first alternative holds, there is nothing to prove. Therefore, throughout this section we assume that the first alternative fails, that is,
\begin{equation}\label{eq:assume}\tag{$\star$}
\text{every two-candidate lottery has distortion larger than }3-\eps,
\end{equation}
and we show that the block described in the second alternative exists.

\subsection{Proof of \texorpdfstring{\Cref{lem:initial}}{the Initial Block Lemma}}

We first restate \cref{lem:initial} to refresh the reader's memory.

\def\abnote{}\initialblock*

By \eqref{eq:assume}, it suffices to show that there is a $\tau_0$-block with mass in $[1/20,3/4]$.

\begin{proof}[of \cref{lem:initial}]
Let $\tau=\tau_0$ and $\xi=\eps$, so that $\xi=\exp(-\tau^{-16})$. Applying \cref{lem:weighted} with $\mu_1=\mu_2$ uniform gives a candidate together with a coupling all of whose edges are valid for it, so $\min_c\sigma(c)=0\le\xi$. Let $a$, $b$, and $p$ be given by \cref{lem:pairtoblock} with $\delta=\sigma$. By \eqref{eq:assume}, the lottery that elects $a$ with probability $p$ and $b$ with probability $1-p$ has distortion larger than $3-\eps$. Hence \cref{lem:pairtoblock} gives a $\tau$-block $(J_0,J_1,J_2)$ and a number $\zeta\in(0,1)$ with properties (i)--(iv). By (iii), the mass of the block is at most $3/4$.

For the lower bound, let $a'\in\{a,b\}$ have the smaller deficiency, and let $\pi$ be a coupling with $\defi_{a'}(\pi)=\sigma(a')$. \Cref{lem:repair}, applied to $\pi$, $a'$, and $J_0$, gives a candidate $c\in J_0$ and a coupling $\pi'$ with
\begin{align*}
\sigma(c)&\le\defi_c(\pi')\\
&\le\sigma(a')+r_{J_0}-t_{J_0}(a') &&\text{(\cref{lem:repair})}\\
&\le\sigma(a')+\frac{48\,(\sigma(a')+\xi)}{\zeta} &&\text{(by \cref{lem:pairtoblock}(ii))}.
\end{align*}
Hence
\begin{align*}
\sigma(c)+\xi&\le\big(\sigma(a')+\xi\big)+\frac{48\,(\sigma(a')+\xi)}{\zeta}\\
&\le\frac{49\,(\sigma(a')+\xi)}{\zeta}.
\end{align*}
By \cref{lem:pairtoblock}(iv), the mass of the block is at least $1/20$.
\end{proof}

\subsection{Proof of \texorpdfstring{\Cref{lem:step}}{the Descent Lemma}}

We now move to \cref{lem:step}, and again restate it to refresh the reader's memory.

\descentstep*

By \eqref{eq:assume}, it suffices to show that for every $\tau_j$-block with mass $s\in[1/20,3/4]$, there is a $\tau_{j+1}$-block with mass $s'\in[1/20,3/4]$ and $s'\le s-1/4000$. Before proving \cref{lem:step}, we fix the setting, construct the new block, and prove five auxiliary claims about it.

\paragraph{Setting.} Fix $0\le j<4000$ and a $\tau$-block $(I_0,I_1,I_2)$ with $\tau=\tau_j$ and mass $s\in[1/20,3/4]$. Let $G=G_{I_0}$ and $\kappa=s^2/4\ge1/1600$. We measure the quality of a candidate by a deficiency that also charges for a loss of mass inside $G$:
\begin{equation}\label{eq:soft}
\delta_I(c)=\min_{\pi}\Big(\defi_c(\pi)+\pos{\kappa-\pi(G,G)}\Big),
\end{equation}
where the minimum is over all couplings. We call $\delta_I$ the \emph{soft protected deficiency}. Clearly $\delta_I(c)\ge\sigma(c)$ for every candidate $c$, so $\delta_I$ satisfies the hypothesis $\delta\ge\sigma$ of \cref{lem:pairtoblock}.

\begin{claim}\label{clm:softmin}
$\min_c\delta_I(c)\le2\tau^2$.
\end{claim}
\begin{proof}
Since $\eps\le\exp(-\tau^{-16})$ and, by \eqref{eq:assume}, no two-candidate lottery has distortion at most $3-\eps$, \cref{lem:protected} gives a candidate $c$ and a coupling $\pi'$ with $\defi_c(\pi')\le2\tau^2$ and $\pi'(G,G)\ge\kappa$. Therefore
\begin{align*}
\delta_I(c)&\le\defi_c(\pi')+\pos{\kappa-\pi'(G,G)}\\
&=\defi_c(\pi')\\
&\le2\tau^2.
\end{align*}
\end{proof}

\paragraph{The new block.} Let $\tau'=\tau_{j+1}$ and $\xi=\max\{\eps,2\tau^2\}$. By the choice of the parameters, 
\begin{align*}
2\tau^2&=2\Big(\tfrac14\exp\big(-\tau'^{-16}\big)\Big)^2\\
&=\tfrac18\exp\big(-2\tau'^{-16}\big)\\
&\le\exp\big(-\tau'^{-16}\big),
\end{align*}so $\eps\le\xi\le\exp(-\tau'^{-16})$, and $\min_c\delta_I(c)\le\xi$ by \cref{clm:softmin}. Let $a$, $b$, and $p$ be given by \cref{lem:pairtoblock} with $\tau'$ in place of $\tau$, the function $\delta_I$, and the scale $\xi$. By \eqref{eq:assume}, the lottery that elects $a$ with probability $p$ and $b$ with probability $1-p$ has distortion larger than $3-\eps$. Hence \cref{lem:pairtoblock} gives a $\tau'$-block $(J_0,J_1,J_2)$ and a number $\zeta\in(0,1)$ with properties (i)--(iv). Write $\hdelta_I=\delta_I+\xi$. By \cref{lem:pairtoblock}(ii), $\hdelta_I(a),\hdelta_I(b)\le\tau'^2$. For $i\in\{a,b\}$, let $\pi_i$ be a coupling that attains the minimum in \eqref{eq:soft}. Since both terms in \eqref{eq:soft} are nonnegative and $\delta_I(i)\le\hdelta_I(i)\le\tau'^2$, we have
\begin{equation}\label{eq:softmass}
\defi_i(\pi_i)\le\delta_I(i)\le\tau'^2\qquad\text{and}\qquad\pi_i(G,G)\ge\kappa-\delta_I(i).
\end{equation}

Let $G'=G_{J_0}$ and $s'=\mu(G')$. Let $a'\in\{a,b\}$ be the candidate with the smaller soft protected deficiency. By \cref{lem:pairtoblock}(ii),
\begin{equation}\label{eq:newcut}
r_{J_0}-t_{J_0}(a')\le48\,\hdelta_I(a')/\zeta\le\tau'^2.
\end{equation}

\paragraph{Auxiliary claims.} We now prove that both candidates $a$ and $b$ lie in the old block, that the new block is nested in the old one, and that the new block has the required mass.

\begin{claim}\label{clm:inside}
Both $a$ and $b$ belong to $I_1$.
\end{claim}
\begin{proof}
Suppose that $i\in\{a,b\}$ is not in $I_1$. Every core voter $v$ of the block ranks $I_0$ above $i$, so every edge from a core voter to a voter of $G$ is invalid for $i$. The core voters cover all of $G$ except for a mass of at most $\tau^2$, so 
\begin{align*}
\defi_i(\pi_i)&\ge\pi_i(G,G)-\tau^2 &&\text{(edges from core voters to $G$ are invalid for $i$)}\\
&\ge\kappa-\delta_I(i)-\tau^2 &&\text{(by \eqref{eq:softmass})}.
\end{align*}
Therefore
\begin{align*}
\kappa&\le\defi_i(\pi_i)+\delta_I(i)+\tau^2\\
&\le2\tau'^2+\tau^2 &&\text{(by \eqref{eq:softmass})}\\
&<1/1600,
\end{align*}
which contradicts $\kappa\ge1/1600$.
\end{proof}

\begin{claim}[Nesting]\label{clm:nest}
$J_0\subseteq I_2$. Consequently, $\mu(G'\setminus G)\le\plu(I_2\setminus I_0)\le2\tau^2\le\xi$.
\end{claim}
\begin{proof}
Suppose that some candidate $j\in J_0$ is not in $I_2$. By \cref{clm:inside}, $a'\in I_1$, so every core voter of the old block ranks $a'$ above $j$, and is therefore not in $T_{J_0}(a')$. Consider the coupling $\pi_{a'}$ and the set $J_0$, and let
\[
\chi^-=\pi_{a'}\big(T_{J_0}(a'),G'\big),\qquad\chi^+=\pi_{a'}\big(V\setminus T_{J_0}(a'),V\setminus G'\big).
\]
Write $\chi^0=\pi_{a'}\big(T_{J_0}(a'),V\setminus G'\big)$. Then
\begin{align*}
r_{J_0}&=\pi_{a'}(V,V\setminus G')=\chi^0+\chi^+,\\
t_{J_0}(a')&=\pi_{a'}\big(T_{J_0}(a'),V\big)=\chi^-+\chi^0,
\end{align*}
and subtracting gives $\chi^+-\chi^-=r_{J_0}-t_{J_0}(a')$. The edges counted in $\chi^-$ are invalid for $a'$, so $\chi^-\le\defi_{a'}(\pi_{a'})\le\tau'^2$ by \eqref{eq:softmass}. Hence
\begin{align*}
\chi^+&=\chi^-+r_{J_0}-t_{J_0}(a')\\
&\le\tau'^2+\tau'^2 &&\text{(bound on $\chi^-$, and \eqref{eq:newcut})}\\
&=2\tau'^2.
\end{align*}
The core voters of the old block are not in $T_{J_0}(a')$, so they send a mass of at most $\chi^+\le2\tau'^2$ to $V\setminus G'$, and the remaining voters of $G$ have mass at most $\tau^2$. Therefore
\begin{align*}
\mu(G\cap G')&\ge\pi_{a'}(G,G\cap G')\\
&=\pi_{a'}(G,G)-\pi_{a'}(G,G\setminus G')\\
&\ge\pi_{a'}(G,G)-\tau^2-2\tau'^2\\
&\ge\kappa-\delta_I(a')-\tau^2-2\tau'^2 &&\text{(by \eqref{eq:softmass})}\\
&\ge\kappa-3\tau'^2-\tau^2 &&\text{(since $\delta_I(a')\le\tau'^2$)}.
\end{align*}
Since $\kappa\ge1/1600$, this is larger than $\tau^2+\tau'^2$, so $G\cap G'$ contains a voter $v$ that is a core voter of both blocks. Since $a'\in I_1$ and $j\notin I_2$, the old block gives $a'\succ_v j$. Since $j\in J_0$ and $a'\notin J_2$ (\cref{lem:pairtoblock}(i)), the new block gives $j\succ_v a'$. This contradiction shows $J_0\subseteq I_2$. Then every voter of $G'\setminus G$ has its top choice in $I_2\setminus I_0$, which proves the second statement.
\end{proof}

\begin{claim}[Mass bounds]\label{clm:mass}
$1/20\le s'\le3/4$.
\end{claim}
\begin{proof}
The upper bound is \cref{lem:pairtoblock}(iii). For the lower bound, apply \cref{lem:repair} to the coupling $\pi_{a'}$ and the set $J_0$. We obtain a candidate $c\in J_0$ and a coupling $\pi'$ with
\[
\defi_c(\pi')\le\defi_{a'}(\pi_{a'})+r_{J_0}-t_{J_0}(a')\qquad\text{and}\qquad\pi'(G,G)\ge\pi_{a'}(G,G)-\mu(G'\setminus G)\ge\pi_{a'}(G,G)-\xi,
\]
where we used \cref{clm:nest}. Since $\pos{\kappa-\pi'(G,G)}\le\pos{\kappa-\pi_{a'}(G,G)}+\xi$, we obtain
\begin{align*}
\delta_I(c)&\le\defi_c(\pi')+\pos{\kappa-\pi'(G,G)}\\
&\le\defi_{a'}(\pi_{a'})+\big(r_{J_0}-t_{J_0}(a')\big)+\pos{\kappa-\pi_{a'}(G,G)}+\xi\\
&=\delta_I(a')+\big(r_{J_0}-t_{J_0}(a')\big)+\xi &&\text{($\pi_{a'}$ attains the minimum in \eqref{eq:soft})}\\
&\le\delta_I(a')+\frac{48\,\hdelta_I(a')}{\zeta}+\xi &&\text{(by \eqref{eq:newcut})}.
\end{align*}
Hence
\begin{align*}
\hdelta_I(c)&=\delta_I(c)+\xi\\
&\le\hdelta_I(a')+\frac{48\,\hdelta_I(a')}{\zeta}+\xi\\
&\le\frac{\hdelta_I(a')}{\zeta}+\frac{48\,\hdelta_I(a')}{\zeta}+\frac{\hdelta_I(a')}{\zeta}\\
&=\frac{50\,\hdelta_I(a')}{\zeta}.
\end{align*}
By \cref{lem:pairtoblock}(iv), $s'\ge1/20$.
\end{proof}

\begin{claim}[Descent]\label{clm:descent}
$s'\le s-s^2/8+4\tau'^2\le s-1/4000$.
\end{claim}
\begin{proof}
Since $a'\notin J_2$, every core voter of the new block ranks $J_0$ above $a'$, so every edge from a core voter of the new block to a voter of $G'$ is invalid for $a'$. Hence
\begin{align*}
\pi_{a'}(G',G')&\le\defi_{a'}(\pi_{a'})+\tau'^2 &&\text{(non-core voters of $G'$ have mass at most $\tau'^2$)}\\
&\le\delta_I(a')+\tau'^2 &&\text{(by \eqref{eq:softmass})}.
\end{align*}
On the other hand, splitting the edges from $G$ to $G$ according to whether their endpoints lie in $G'$,
\begin{align*}
\kappa-\delta_I(a')&\le\pi_{a'}(G,G) &&\text{(by \eqref{eq:softmass})}\\
&\le\pi_{a'}(G',G')+\pi_{a'}(G\setminus G',V)+\pi_{a'}(V,G\setminus G')\\
&=\pi_{a'}(G',G')+2\mu(G\setminus G').
\end{align*}
Moreover,
\begin{align*}
\mu(G\setminus G')&=\mu(G)-\mu(G')+\mu(G'\setminus G)\\
&=s-s'+\mu(G'\setminus G)\\
&\le s-s'+2\tau^2 &&\text{(\cref{clm:nest})}.
\end{align*}
Combining the three inequalities,
\begin{align*}
\kappa-\delta_I(a')&\le\delta_I(a')+\tau'^2+2\big(s-s'+2\tau^2\big),
\end{align*}
and therefore
\begin{align*}
2s'&\le2s-\kappa+2\delta_I(a')+\tau'^2+4\tau^2\\
&\le2s-\kappa+3\tau'^2+4\tau^2 &&\text{(since $\delta_I(a')\le\tau'^2$)}\\
&\le2s-\kappa+7\tau'^2.
\end{align*}
Dividing by two and using $\kappa=s^2/4$,
\begin{align*}
s'&\le s-\frac{s^2}{8}+\frac{7\tau'^2}{2}\\
&\le s-\frac{s^2}{8}+4\tau'^2\\
&\le s-\frac{1}{3200}+4\tau'^2 &&\text{(since $s\ge1/20$)}\\
&\le s-\frac{1}{4000}.
\end{align*}
\end{proof}

Finally, we combine the auxiliary claims to prove \cref{lem:step}.

\begin{proof}[of \cref{lem:step}]
Let $(I_0,I_1,I_2)$ be a $\tau_j$-block with mass $s\in[1/20,3/4]$, and let $(J_0,J_1,J_2)$ be the $\tau_{j+1}$-block constructed above, with mass $s'$. By \cref{clm:mass}, $s'\in[1/20,3/4]$, and by \cref{clm:descent}, $s'\le s-1/4000$. Hence the second alternative of \cref{lem:step} holds.
\end{proof}

%% file: src/computation.tex
\section{Computing the Lottery}\label{sec:computation}

The proof of \cref{thm:main} in \cref{sec:overview} shows that a good two-candidate lottery exists, but it does not describe it explicitly. In this section, we show that a two-candidate lottery of minimum distortion can be found in polynomial time by linear programming. In particular, by the existence part of \cref{thm:main}, the lottery that we compute has distortion at most $3-\eps_*$.

Fix two candidates $a,b$ (possibly equal) and a reference candidate $o$. Let $\mathcal M_o$ be the polyhedron of all functions $d$ on pairs of points of $V\cup C$ that are symmetric, nonnegative, vanish on the diagonal, satisfy the triangle inequality and the consistency constraints $d(v,c)\le d(v,j)$ for $c\succ_v j$, and satisfy the normalization $\SC_d(o)=1$. This polyhedron has $O((n+m)^2)$ variables and $O((n+m)^3)$ constraints, and it is nonempty (all off-diagonal distances equal to one is a point of it). For $p\in[0,1]$, define
\[
f_o(p)=\sup_{d\in\mathcal M_o}\big(p\,\SC_d(a)+(1-p)\,\SC_d(b)\big).
\]

\begin{observation}\label{obs:lpdist}
The distortion of the lottery that elects $a$ with probability $p$ and $b$ with probability $1-p$ is $\max_{o\in C}f_o(p)$.
\end{observation}
\begin{proof}
Let $\rho$ be the distortion. Every consistent metric with $\OPT>0$ can be scaled so that an optimal candidate $o$ has cost one, and then it belongs to $\mathcal M_o$; hence $\max_of_o(p)\ge\rho$. Conversely, every $d\in\mathcal M_o$ satisfies 
\begin{align*}
p\,\SC_d(a)+(1-p)\SC_d(b)&\le\rho\,\OPT_d\\
&\le\rho\,\SC_d(o)\\
&=\rho,
\end{align*}
and therefore $\max_of_o(p)\le\rho$. Here the first inequality also holds when $\OPT_d=0$, by \cref{sec:prelim}.
\end{proof}

For fixed $p$, the value $f_o(p)$ is the optimum of a linear program whose objective depends linearly on $p$ and whose constraints do not depend on $p$. By linear programming duality, $f_o(p)\le \rho'$ if and only if the dual program has a feasible solution with objective value at most $\rho'$ (if the dual is infeasible, the primal is unbounded and $f_o(p)=\infty$). The dual constraints are linear in the dual variables and in $p$ jointly, because $p$ only appears in the right-hand side of the dual constraints. Therefore, the problem of minimizing $\rho'$ subject to $p\in[0,1]$ and, for every reference candidate $o$, a feasible dual solution of value at most $\rho'$, is a single linear program of polynomial size. Its optimal value is the smallest distortion of a lottery supported on $\{a,b\}$. Solving this program for each of the $O(m^2)$ pairs and taking the best pair gives a two-candidate lottery of minimum distortion in polynomial time. This proves the algorithmic part of \cref{thm:main}.

%% file: src/conclusion.tex
\section{Conclusion and Open Problems}\label{sec:conclusion}

We showed that the metric distortion barrier of $3$ can be broken by randomizing over only two candidates, if the two candidates may receive unequal probabilities. Our proof generalizes the warm-up of \cref{sec:warmup}, which relies on a single popular candidate, by replacing this candidate with a popular group of candidates, formalized as a fuzzy block. The main theorem then follows from two lemmas: an initial block exists, and every block can be replaced by a block of noticeably smaller mass. Both lemmas follow from four building blocks: the weighted veto lemma, the repair lemma, the pair-to-block lemma, and the protected matching lemma. We also showed that a simple explicit two-candidate lottery achieves distortion $3-2r^2/(1-r+r^2)$ when some candidate is the top choice of an $r$ fraction of the voters. We conclude with a few open problems.

\begin{itemize}
\item The most natural question is the conjecture of Cai et al.~\cite{bounded}: is there a rule that always selects two candidates and elects each of them with probability $1/2$, and has distortion $3-\Omega(1)$? The lotteries used in our proof either weigh the two candidates inversely proportionally to their deficiencies or take their weights from a coupled veto process, and in both cases the weights may be very far from uniform.
\item The constant $\eps_*$ in \cref{thm:main} is extremely small. What is the smallest distortion that two-candidate lotteries can guarantee? A meaningful numerical bound would likely require a more efficient certificate, or a direct characterization of the optimal two-candidate lottery.
\item Our lottery is computed by a linear program. Is there a simple and explicit two-candidate rule, in the spirit of Plurality Veto, with distortion $3-\Omega(1)$? \Cref{thm:plurality} gives such a rule when some candidate has a large plurality share, and \cref{ex:star} shows that in general the rule must sometimes elect candidates that are no voter's top choice.
\end{itemize}

%% file: src/ai.tex
\section*{Use of AI}

We obtained the results of this paper through interactive discussions with ChatGPT and Claude. All proofs are given in full in the paper. The authors have verified all proofs and take full responsibility for the correctness of the results.